\documentclass{aa} 

\usepackage[utf8]{inputenc}
\usepackage{graphicx}
\usepackage{txfonts}
\usepackage{aas_macros}
\usepackage{hyperref} 
\hypersetup{colorlinks,citecolor=blue,linkcolor=blue,urlcolor=blue}
\usepackage{ulem}
\def\msun{M$_\odot$}
\def\lsun{L$_\odot$}

\newcommand{\iso}[1]{$^{#1}$}

\begin{document} 

   \title{Nucleosynthetic yields of ${\rm Z=10^{-5}}$ intermediate-mass stars}

   \subtitle{}

   \author{P. Gil-Pons
          \inst{1,2}
          \and
          C.L. Doherty \inst{3,4} \and J. Guti\'errez \inst{1,2}
          \and
          S. W. Campbell\inst{4} \and
          L. Siess\inst{5} \and J. C. Lattanzio\inst{4} \\
          }

   \institute{EETAC, Universitat Polit\`ecnica de Catalunya, Campus Baix Llobregat, C3, 08840 Castelldefels, Spain.\\
        \email{pilar.gil@upc.edu}
         \and
             Institut d'Estudis Espacials de Catalunya, Ed- Nexus Campus Nord, Barcelona, Spain.
        \and
            Konkoly Observatory, Hungarian Academy of Sciences, 1121 Budapest.
        \and 
        School of Physics and Astronomy, Monash University, Australia.
       \and Institut d'Astronomie et d'Astrophysique, Universit\'e Libre de Bruxelles, ULB, Belgium%
             }

\date{Received Month Day, Year; accepted Month Day, Year}

 
  \abstract
   {Observed abundances of extremely metal-poor stars in the Galactic Halo hold
   clues for the understanding of the ancient universe. Interpreting these clues
   requires theoretical stellar models in a wide range of masses at the
   low-metallicity regime. The existing literature is relatively rich with
   extremely metal-poor massive and low-mass stellar models. However, the
   evolution of intermediate-mass stars of ${\rm Z \lesssim 10^{-5}}$ remains
   poorly known, and the impact of the uncertain input physics on the evolution
   and nucleosynthesis has not been systematically analysed.}
   {We aim to provide the nucleosynthetic yields of intermediate-mass ${\rm Z =
   10^{-5}}$ stars between 3 and 7.5 \msun, and quantify the effects of the
   uncertain wind rates. We expect these yields can be eventually used to assess
   the contribution to the chemical inventory of the early universe, and to help
   interpret abundances of selected C-enhanced extremely metal-poor stars (CEMP)
   stars. }
   {We compute and analyse the evolution of surface abundances and
   nucleosynthetic yields of ${\rm Z = 10^{-5}}$ intermediate-mass stars, from
   their main sequence till the late stages of their thermally-pulsing (Super)
   AGB phase, with different prescriptions for stellar winds. We use the
   postprocessing code \textsc{MONSOON} to compute the nucleosynthesis based on
   the evolution structure obtained with the Monash-Mount Stromlo stellar
   evolution code \textsc{MONSTAR}.
   By comparing our models and other existing in the literature, we explore
   evolutionary and nucleosynthetic trends with wind prescriptions and with
   initial metallicity (in the very low-Z regime). We also compare our
   nucleosynthetic yields to observations of CEMP-s stars belonging to the
   Galactic Halo.}
   {The yields of intermediate-mass extremely metal-poor stars reflect the
   effects of very deep or corrosive second dredge-up (for the most massive
   models), superimposed with the combined signatures of hot-bottom burning and
   third dredge-up. Specifically, we confirm the reported trend that models with
   initial metallicity ${\rm Z_\mathrm{ini} \lesssim 10^{-3}}$ give positive
   yields of ${\rm ^{12}}$C, ${\rm ^{15}}$N, ${\rm ^{16}}$O, and ${\rm
   ^{26}}$Mg. The ${\rm ^{20}}$Ne, ${\rm ^{21}}$Ne, and ${\rm ^{24}}$Mg yields,
   which were reported to be negative at ${\rm Z_\mathrm{ini} \gtrsim 10^{-4}}$,
   become positive for ${\rm Z = 10^{-5}}$. The results using two different
   prescriptions for mass-loss rates differ widely in terms of the duration of
   the thermally-pulsing (Super) AGB phase, overall efficiency of the third
   dredge-up episode, and nucleosynthetic yields. We find that the most
   efficient of the standard wind rates frequently used in the literature seems
   to favour agreement between our yield results and observational data.
   Regardless of  the wind prescription, all our models become N-enhanced EMP
   stars. }
  {}

 \keywords{
    nuclear reactions, nucleosynthesis, abundances--stars: evolution -- stars: Population II -- stars: AGB and post-AGB -- ISM: abundances.
    }

\maketitle
%
\section{Introduction}

Stellar evolution models and nucleosynthetic yields of extremely metal-poor
stars (hereafter EMPs) are important pieces of the big puzzle of the chemical
evolution history of the universe.  EMPs are those with ${\rm Z \lesssim
10^{-5}}$ or [Fe/H]$\lesssim -3$ \citep{christlieb2005} \footnote{Metallicity
may be expressed as referred to the solar value according to the standard
expression $\mathrm [Fe/H]=\log{(N_{Fe}/N_H)_\star}-\log{(N_{Fe}/N_H)_{\odot}}$. 
Abundances of an elementor isotope X can also expressed as referred to solar as 
$\mathrm [X/Fe]=\log{(N_{x}/N_{Fe})_\star}-\log{(N_{X}/N_{Fe})_{\odot}}$}. 
A number of interesting works have been presented in this field during the last decades
(e.g.  \citealt{fujimoto1984}, \citealt{cassisi1993}, \citealt{fujimoto2000},
\citealt{marigoetal2001}, \citealt{chieffi2001}, \citealt{siess2002},
\citealt{denissenkov2003}, \citealt{herwig2014}, \citealt{gil07a},
\citealt{lau2009}, \citealt{campbell2008}, \citealt{suda2010} and
\citealt{gilpons2013}, to mention a few). However, the evolution,
nucleosynthesis, and even the fates of intermediate-mass stars between $\sim$ 3
and $\sim$ 7-9 \msun{} is still poorly constrained, and theoretical results from
different authors frequently differ widely \citep{gilpons2018}.  

The above mentioned works not withstanding, the amount of results reported in
this metallicity regime is much smaller than those dedicated to
intermediate-mass stars of higher metallicity (see the review by
\citealt{karakas2014}, and references therein). The main reasons are, probably,
the historical difficulty in the detection of stars at the lowest metallicities;
the very substantial uncertainties which hamper our knowledge of these stars
(mainly related to the treatment of convection, mixing and mass-loss rates,
which can hardly be calibrated by observations); the fact that evolutionary
calculations of these objects tend to involve the computation of huge numbers of
thermal pulses \citep{lau2008}; and finally the limitations of 1-dimensional
hydrostatic codes for computing certain phases of the evolution of
low-metallicity stars \citep{woodward2015}, such as dual-shell flashes
\citep{campbell2008, mocak2010}.  These limitations in our knowledge of the
evolution and fates of intermediate mass EMPs (IM EMPs) necessarily imply an
even poorer understanding of the associated nucleosynthesis, and thus our
knowledge of their yields is restricted to a limited number of isotopes (see,
e.g. \citealt{abia2001}, \citealt{iwamoto2009}, and \citealt{gilpons2013}). 

The lack of nucleosynthetic yields of intermediate-mass stars at the lowest
Z-regime ($Z\lesssim 10^{-5}$) is also reflected in the inputs of chemical
evolution models (see, for instance, \citealt{chiappini1997},
\citealt{kobayashi2006}, \citealt{tsujimoto2012}, \citealt{brusadin2013},
\citealt{molla2015}, \citealt{cote2016}, \citealt{matteucci2016},
\citealt{spitoni2017}, \citealt{prantzos2018}, \citealt{millan2019} and
references therein).  Frequently used sets of yields, such as
\cite{vandenhoek1997}, \cite{marigo2001}, \cite{gavilan2005}, or
\cite{gavilan2006} are based on a synthetic approach for the treatment of the
thermally-pulsing Asymptotic Giant Branch, TP-AGB.
\citet{ventura2005,ventura2010}, \cite{siess2010}, \citet{karakas2010},
\citet{doherty2014a}, \citet{doherty2014b} and \citet{ritter2018} presented
detailed evolution and nucleosynthetic yield calculations of AGB and super-AGB
stars\footnote{Super-AGB stars are those which undergo a C-burning phase prior
to the thermally pulsing phase, the TP-(S)AGB - for a recent review see
\citealt{doherty2017}.}, but they did not study models below $Z=10^{-4}$.
\citet{campbell2008} presented primordial to extremely metal-poor stars yields
up to 3\msun. \citet{iwamoto2009} computed the evolution and nucleosynthesis of
intermediate-mass models at Z=2$\times$10$^{-5}$.
\citet{cristallo2009,cristallo2011,cristallo2015} presented yields for low and
intermediate mass stars in the metallicity range (-2.15 $\leq$ [Fe/H] $\leq$
+0.15). \citet{chieffi2001} and \cite{abia2001} computed yields of primordial
stars in a wide mass range (between low and massive). The evolution and
nucleosynthesis in the poorly-explored lowest-metallicity regime involve
peculiar phenomena, such as double-shell and double-core flashes
\citep{campbell2008}, the corrosive second dredge-up (\citealt{gilpons2013},
\citealt{doherty2014b}) and the hot-third dredge-up (\citealt{chieffi2001},
\citealt{goriely2004}, \citealt{herwig2004}, \citealt{gilpons2013},
\citealt{straniero2014}), which definitely deserve a proper analysis. 

From the point of view of observations of metal-poor stars, relevant for the
understanding of the evolution of the primitive universe, it is important to
recall the amount of data from big surveys, such as the HK objective-prism
survey \citep{beers1992}, the Hamburg-ESO survey \citep{christlieb2002},
SkyMapper \citep{keller2007}, the Sloan Extension for Galactic Understanding and
Exploration, SEGUE \citep{yanny2009}, and the  Large Sky Area Multi-Object Fibre
Spectroscopic Telescope, LAMOST \citep{cui2012} to be further expanded as WEAVE
\citep{dalton2012}, the PRISTINE survey \citep{starkenburg2014}, and,
especially, with the James Webb Space Telescope \citep{zackrisson2011}. At
present almost 1000 stars with [Fe/H]$\lesssim$-3 have been detected in the
Milky Way and dwarf galaxies. They tend to show high abundance dispersion, which
reflects stochasticity in the nature of their progenitors: one CEMP star may
show the signature of a single or very few stars, which makes them excellent
laboratories to test early stellar nucleosynthesis.   

This work is aimed to address the problem of the scarcity of nucleosynthetic
yield calculations in the EMP regime, to explore the effects of input physics,
and to compare our theoretical results with observational data in the relevant
metallicity regime. We present the yields of ${\rm Z = 10^{-5}}$ stars of
initial masses between 3 and 7.5 \msun{}, slightly overlapping the upper mass
limit presented in \citet{campbell2008} ($\leq$3\msun{}).  This manuscript also
represents an extension of the work by \citet{doherty2014a,doherty2014b}, in
which the nucleosynthesis of massive AGB and Super-AGB stars from ${\rm
Z=10^{-4}}$ to solar metallicity (Z=0.02) was analysed. Part of the stellar
structure calculations upon which this work is based were presented in
\citet{gilpons2013}. In order to quantify the effect of the uncertain wind
mass-loss rates, we include new models computed with an alternative mass-loss
prescription, and postprocess all the stellar structure results with a
nucleosynthesis code. We also explore the main trends of evolution and yields
with metallicity in the metal-poor regime.

The results presented have potential interest in the context of stellar
archaeology (see e.g. \citealt{frebel2015}), because comparison between
theoretical gas yields and observations of CEMP stars in the Halo and dwarf
galaxies may help to gain insight into the chemistry of the early universe, and
into the nature of the early stellar populations.  Such comparisons can also
help constraining the uncertain input physics of CEMP stars, specifically, those
related to the efficiency of stellar winds.
They could help understand the pollution history of the intracluster medium and,
eventually, the origin of multiple populations in globular clusters
\citep{ventura2016c}.  Used as inputs for Galactic chemical evolution (GCE)
models, our yields can also help to understand the chemical evolution of the
early universe, and to probe the Milky Way structure and formation history
\citep{gibson2003}.
Our results can also be used as input physics for population synthesis, to help
constrain the primitive IMF \citep{suda2013}, and understand mass transfer in
low and intermediate-mass metal-poor stars \citep{abate2016}.

This manuscript is organised as follows. Section \ref{sec:evol} summarises the
evolution of ${\rm Z=10^{-5}}$ intermediate-mass EMP stars (IM EMPs), and
describes trends with different wind prescriptions and metallicities. Section
\ref{sec:evonuc} analyses the nucleosynthesis and the evolution of surface
abundances of our computed models. Section \ref{sec:yields} presents our yields
and production factors, Section \ref{sec:chemev} explores the trends with
metallicity and input physics of yields computed here and other existing in the
literature, and Section \ref{sec:binary} presents a brief discussion of the
binary channel for the formation of EMP stars, and a comparison to observational
data. Finally, Section \ref{sec:finale} summarises and draws the main
conclusions from our work.


\begin{table*}[t]
    \caption{Main characteristics of the TP-(S)AGB of our Z=10$^{-5}$ models.
	$M_\mathrm{ini}$ corresponds to the initial mass. $N_{TP}$,
	$\tau_{TP-(S)AGB}$ and $\Delta t_{IP}$ are, respectively, the number of
	thermal pulses, the duration of the TP-(S)AGB (given from the first
	thermal pulse until the end of our computations), and the interpulse
	period.  $M_{c,ini}$, $N_{c,f}$ and $M_{env,f}$ are, respectively, the
	masses of the H-exhausted cores prior to the TP-(S)AGB, and the masses
	of the H-exhausted cores and the remnant envelopes at the end of the
	TP-(S)AGB. 
        $\langle T_{HeBS}\rangle$, $\langle T_{HBS}\rangle$,  $\langle
	T_{BCE}\rangle$ are, respectively, the temperatures at the times of
	maximum luminosity for each pulse, given at the centre of the He-burning
	shell, at the H-burning shell and at the base of the convective envelope
	respectively, and averaged over the number of thermal pulses in each
	sequence $M_{dredge}^{tot}$ is the total mass dredged-up.
	$\langle\lambda \rangle$ was also averaged over the number of thermal
	pulses in each case, and $\langle\dot M_{wind}\rangle$ is the average
	mass-loss rates due to winds, that is, the envelope mass lost over the
	duration of the (S)AGB phase. 
        Models were calculated using both the \citet{vassiliadis1993} and
	\citet{bloecker1995} with $\eta$=0.02 mass-loss rate prescription. The 6
	\msun{} model includes additional calculations with $\eta$=0.04 and
	$\eta$=1. Some of the entries for models between 4 and 7 \msun, computed
	with VW93, were presented in Tables 1 and 3 of \citet{gilpons2013}. We
	show them here to facilitate comparison.   }
    \centering
    \begin{tabular}{lcccccccccccc}
    \hline
    \hline
    $M_\mathrm{ini}$ & $N_{TP}$ & $\tau_{TP-(S)AGB}$ 
    & $\Delta t_{IP}$ & $M_{c,ini}$ & $M_{c,f}$ & $M_{env,f}$ & $\langle T_{HeBS}\rangle$ & 
    $\langle T_{HBS}\rangle$ & $\langle 
    T_{BCE}\rangle$ & $M_{dredge}^{tot}$ & 
    $\langle\lambda\rangle$ & $\langle\dot 
    M_{wind}\rangle$\\[1pt]
    \msun & & Myr & yr & \msun & \msun & \msun & $MK$ & $MK$ & $MK$ &  \msun& &\msun/yr\\[1pt]
    \hline \\[-4pt]    
    \multicolumn{13}{c}{{\bf VW93}}\\[1pt]
       $3$  &  122  & 1.41 & 16148 & 0.81 & 0.87 & 0.10 & 345  & 83  & 68 &   0.70 & 0.98 & 4.0$\times 10^{-7}$\\
       $4$  & 197 & 1.19 & 6702 & 0.87 & 0.93 & 0.25 & 359 & 91 & 85 & 0.53 & 0.92 & 1.9$\times 10^{-6}$ \\
       $5$  & 213 & 0.95 & 4382 & 0.91 & 0.97 & 0.54 & 357 & 95 & 88 & 0.40 & 0.91 & 3.7$\times 10^{-6}$ \\
       $6$  & 372 & 0.87 & 1828 & 0.98 & 1.04 & 0.69 & 362 & 104 & 99 & 0.36 & 0.85 & 6.2$\times 10^{-6}$ \\
       $7$  & 607 & 0.50 & 739 & 1.05 & 1.14 & 0.73 & 367 & 117 & 114 & 0.24 & 0.78 & 1.1$\times 10^{-5}$ \\
       \hline \\[-4pt]
        \multicolumn{13}{c}{{\bf Blo95}}\\[1pt]
        $3$  & 23 & 0.45 & 17750 & 0.81 & 0.84 & 0.54 & 320  & 88  & 34 &   0.15 & 0.80 & 3.2$\times 10^{-6}$\\
        $4$  & 35 & 0.31 & 8285 & 0.87 & 0.89 & 0.62 & 328  & 95  & 62 &   0.14 & 0.88 & 8.3$\times 10^{-6}$\\
        $5$  & 43 & 0.26 & 5681 & 0.91 & 0.93 & 0.01 & 329  & 101  & 73 &   0.12 & 0.78 & 1.6$\times 10^{-6}$\\
        $6$  & 60 & 0.17 & 2574 & 0.96 & 0.98 & 0.20 & 330  & 110  & 89 &   0.09 & 0.72 & 2.8$\times 10^{-5}$\\
        $6\:({\eta=0.04})$ & 41 & 0.13 & 2523 & 0.96 & 0.98 & 0.01 &319 & 112 & 84 &   0.002  & 0.66 &  3.9$\times 10^{-5}$ \\
        $6\:({\eta=1})$ & 16 & 0.03 & 2393  & 0.96 & 0.97 & 0.01 & 283 & 111 & 54    & 0.001 & 0.33 & 2.0$\times 10^{-4}$\\
        $7$  & 82 & 0.11 & 1428  & 1.05 & 1.07 & 0.40 & 325 & 126 & 117 & 0.015 & 0.55 & 5.6$\times 10^{-5}$ \\
        $7.5$  & 254 & 0.06 & 423 & 1.13 & 1.16 & 0.42 & 300 & 146 & 142 & 0.005 & 0.01 & 9.1$\times 10^{-5}$ \\
       \\
    \end{tabular}
    \label{tab:evol}
\end{table*}

\section{Summary of the structural evolution of intermediate-mass EMP models} \label{sec:evol}

We now briefly present the code and summarize the main results presented in
\citet{gilpons2013} in order to compare with the new calculations presented in
this work. This new release includes a range of new models of initial masses
between 3 and 7.5 \msun, computed with an alternative wind prescription during
the AGB phase. We also intend to emphasize here the main characteristics of the
evolution which ultimately determine the nucleosynthesis described in detail in
the forthcoming sections.

\subsection{Code and input physics description}\label{sec:code}

Our nucleosynthetic calculations are the result of postprocessing on existing
structure profiles computed with the Monash-Mount Stromlo code \textsc{MONSTAR}
(\citealt{frost1996}, \citealt{campbell2008}), and presented in
\cite{gilpons2013} and \citet{gilpons2018}. Its main characteristics  involve
the determination of convective boundaries using the Schwarzschild criterion,
complemented with the attempt to search for neutrality approach (see, e.g.
\citealt{castellani1971}, \citealt{frost1996}). The mixing-length to pressure
scale height ratio   $\alpha$ was taken as 1.75. 

\subsubsection{Mass loss}

Mass-loss rates during the red giant branch are set to follow the prescription
by \citet{reimers1975}. Note however that mass loss in this part of the
evolution is irrelevant at the considered initial metallicity (see Section
\ref{sec:evoprev}). Mass-loss rates during the AGB or Super-AGB phase follow
\cite{vassiliadis1993} (VW93 henceforth). These authors determined a direct
relation between mass-loss rate and pulsation period from an analysis of CO
microwave observations of AGB stars.  To test the dependence of our results on
this physical input, additional calculations are presented in this work,
following \citet{bloecker1995} (Blo95 henceforth). Blo95 formulation is based on
the \cite{reimers1975} mass-loss rate, and considers atmospheric calculations
for Mira stars made by \cite{bowen1988}. It includes a parameter $\eta$ to be
determined by calibration. For instance \citet{ventura2000} proposed a value
$\eta=0.02$ based on observations of Li-rich giants in the Large Magellanic
Cloud. This value for $\eta$ was also used, for instance, in \citet{ventura2001}
and \citet{doherty2014b}. 
VW93 is the standard mass-loss prescription when MONSTAR (and similar versions
of this code) is used. However, \citet{reimers1975} with $\eta$ values between 5
and 10 was used by \citet{karakas2010} to compute $Z=10^{-4}$ models, and Blo95
with $\eta=0.02$ is more frequently considered by authors computing very low-Z
models (e.g. \citealt{herwig2004}).
 
It is also commonly assumed that the mass loss rate depends on the metallicity.
\cite{kudritzi1989} proposed a scaling relation of the form:

\begin{equation}
    \dot{M}(Z_\mathrm{surf}) = \left( \dfrac{Z_\mathrm{surf}}{Z_\odot}\right)^n  \dot{M}(Z_\odot) 
\end{equation}

\noindent where $Z_\mathrm{surf}$ is the stellar metal surface abundance, and
$n$ is an exponent typically ranging between 0.5 and 0.7. However, the arguments
for this scaling relation were derived from line-driven winds, and thus relevant
for more massive and hotter stars than the ones we are considering. The
characteristic higher luminosity of low-Z stars, their surface C-abundances,
pulsations, and their possibility to form dust might allow for wind mechanisms
(and rates) not so different from those of higher Z stars (e.g.
\citealt{mattsson2008}, \citealt{lagadec2008}). Therefore, we opted for not
introducing any Z-scaling.

\subsubsection{Opacities and Initial composition}
Interior stellar opacities are from \citet{igl96}. We use low-temperature
opacities that take into account composition changes in C, N, and O. It has been
shown that this is critical in this metallicity range, both for the duration of
the TP-AGB phase, and for the overall efficiency of nucleosynthesis and mixing
during this stage \citep[see][]{constantino2014}. Low-temperature opacity tables
are from \citet{lederer2009} and \citet{marigo2009}.\\ 

Initial composition was solar scaled as in \cite{gre93}.  EMP stars are known to
have $\alpha$-enhancements, however we opted for keeping continuity with the
existing grid in \citet{doherty2014a,doherty2014b} and defer $\alpha$-enhanced
calculations for a future work. 
Only the isotopes relevant for the energy generation were considered in the
stellar structure calculations ($^1$H, $^{3}$He, $^{4}$He, $^{12}$C, $^{14}$N,
$^{16}$O, and Z$_{other}$, which included all other species). Nuclear reaction
rates were taken from \citet{caughlan1988}, and updated with NACRE
\citep{angulo1999}.


\begin{figure*}
\begin{center}
\includegraphics[width=1\linewidth]{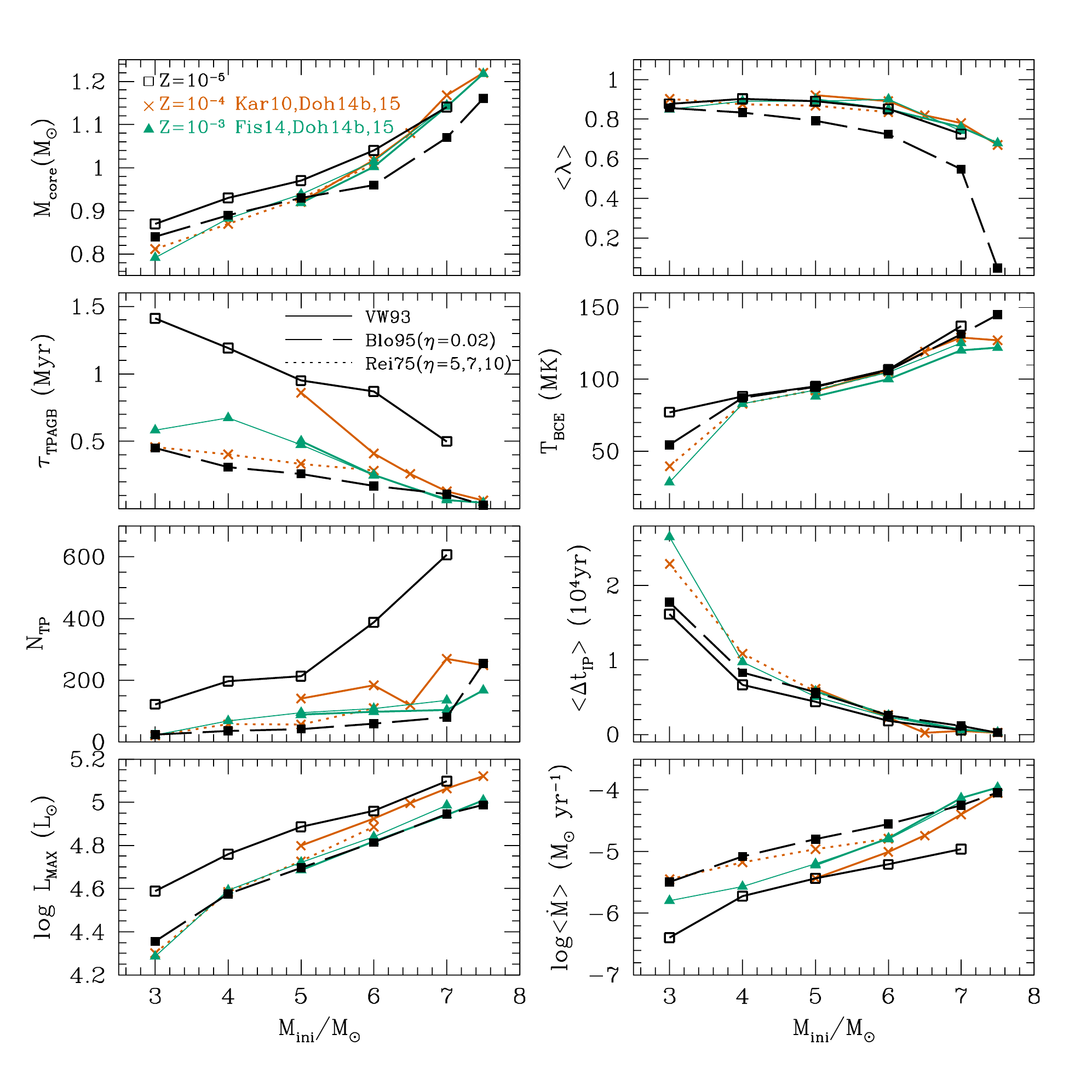}
\caption{Relevant TP-(S)AGB parameters for intermediate-mass models of different
	metallicities calculated with similar versions of the Monash-Mount
	Stromlo stellar evolution code.  $M_\mathrm{core}$ is the core mass at
	the end of the TP-(S)AGB calculations, $\lambda$ is the average TDU
	parameter, $\tau_{TPAGB}$ is the duration of the TP-(S)AGB phase,
	$T_{BCE}$ is the maximum temperature at the base of the convective
	envelope, $N_{TP}$ is the number of thermal pulses, $t_{IP}$ is the
	average interpulse period, $L_{MAX}$ is the maximum luminosity and
	$\dot{M}$ is the average mass-loss rate during the TP-(S)AGB phase.
	Values are shown for our $Z=10^{-5}$ models (black); for $Z=10^{-4}$
	models from \citet{karakas2010} (dotted orange), and from
	\citet{doherty2014b, doherty2015} (solid orange); for $Z=10^{-3}$ models
	for \citet{fishlock2014} (thin green) and \citet{doherty2014b,
	doherty2015} (thick green).  Sequences calculated with the wind
	prescriptions by VW93, \citet{reimers1975}, and Blo95 are shown,
	respectively, with solid, dotted and dashed lines. 
}
\label{fig:panelevo}
\end{center}
\end{figure*}

\begin{figure*}[ht]
\begin{center}
\vspace{-0.2cm}
\includegraphics[width=0.8\linewidth]{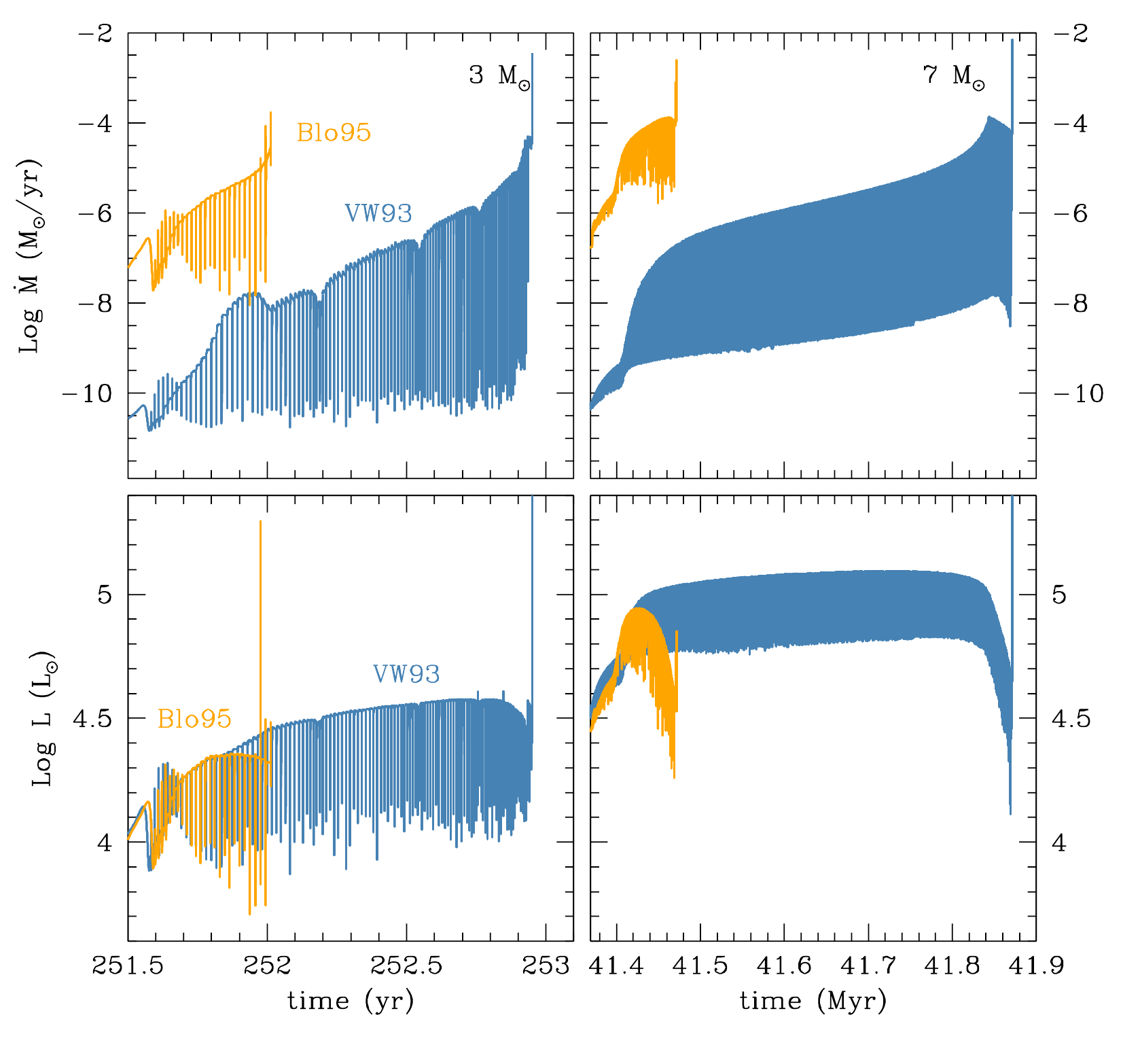}
\caption{Left panels: evolution of mass-loss rates (upper) and total
	luminosities (lower) for the 3~\msun{} $Z=10^{-5}$ model computed with
	the prescriptions of \citet{vassiliadis1993} (blue) and
	\citet{bloecker1995} with $\eta=0.02$ (orange). Right panel: same for
	the 7 \msun{} $Z=10^{-5}$ models.} \label{fig:mloss}
\end{center}
\end{figure*}

\subsection{Evolution up to the TP-(S)AGB} \label{sec:evoprev}

The details of the evolution of intermediate-mass $Z=10^{-5}$ stars were
thoroughly described in \citet{gilpons2013}. Here we present a summary of the
most relevant features for the nucleosynthesis, as well as additional sequences
calculated both with the wind rate prescriptions by VW93, as in
\citet{gilpons2013}, and by Blo95 with parameter $\eta$=0.02, 0.04 and 1. 

Our models experienced a pre-TP-(S)AGB evolution characteristic of IM EMPs of
Z$\lesssim$10$^{-5}$, with core H- and He-burning occurring before the first
ascent of the giant branch and its associated dredge-up episode
\citep{gir96,chieffi2001}.  Our 7 and 7.5 \msun{} models experience off-centre
C-burning and eventually develop degenerate ONe cores surrounded by CO shells. 
This process is well known since the 1990s (see, for instance,
\citealt{gbiben1994}, \citealt{siess2006}, \citealt{gilpons2003},
\citealt{doherty2010}, \citealt{jones2013})

Models of initial mass M$_\mathrm{ini}$ between 3 and 6.5 \msun{} experience a
standard second dredge-up (SDU) episode during which the convective envelope
advances inwards and reaches regions of the star previously processed by
H-burning via the CN-cycle. More massive models undergo the corrosive SDU in
which the base of the convective envelope advances even deeper, and reaches
regions processed by He-burning.  Given that our models do not include rotation,
the first event to alter the surface composition is the (corrosive) SDU.  
Recall that stars of metallicity ${\rm Z=10^{-5}}$ do not experience a first
dredge-up episode, but we keep the nomenclature SDU to refer to the dredge-up
occurring during the Early-AGB, for consistency with the evolution of higher
metallicity stars.


\subsection{Evolution during the TP-(S)AGB}\label{sec:evotp}

Metallicity influences the evolution of the TP-(S)AGB of intermediate-mass stars
mainly through its effects on mass-loss rates and thus, on the duration of this
phase. Figure \ref{fig:panelevo} shows a summary of relevant parameters during
the TP-(S)AGB stars of our $Z=10^{-5}$ sequences, compared to $Z=10^{-4}$ and
$Z=0.001$ sequences which were computed with similar versions of
\textsc{monstar} (see Section \ref{sec:code}). Note that comparisons are not
straightforwards, as the input physics in these versions of the code present
some relevant variations. VW93 mass-loss rates are used in all the cases, except
our $Z=10^{-5}$ models calculated with Blo95, and the $Z=10^{-4}$ sequences by
\citet{karakas2010}, who used the prescription by \citet{reimers1975}, corrected
with a multiplying parameter $\eta_R$ varying between 5 (for the 3 \msun{}
model), 7 (for the 4 \msun{} model) and 10 (for models of $M_\mathrm{ini}\geq
5\:$\msun).  \citet{karakas2010} also scaled mass-loss rates with metallicity as
$\dot M=\sqrt{Z/Z_{\odot}}\: \eta_R \dot M_R$, where $\dot M_R$ represents the
mass-loss rates exactly as in \citet{reimers1975}. As we will see, these
differences in wind prescriptions strongly affect the TP-AGB evolution and,
ultimately, the nucleosynthetic yields. 

\subsubsection{Third dredge-up}

All our models experience third dredge-up (TDU) to varying extents. The
efficiency of this process is described by the parameter $\lambda=\frac{\Delta
M_\mathrm{dredge}}{\Delta M_\mathrm{core}}$, where $\Delta M_\mathrm{dredge}$ is
the H-exhausted core mass dredged-up by the convective envelope after a thermal
pulse, and $\Delta M_\mathrm{core}$ is the amount by which the core has grown
during the previous interpulse period.  $\lambda$ tends to increase during the
first 10s thermal pulses, and remain almost constant during the remaining
TP-(S)AGB.
This parameter is known to decrease with increasing mass \citep{straniero2003}.
The reason is ascribed to the fact that more massive AGB stars have hotter and
more compact cores. The temperature in the He-burning shell is higher, radiation
pressure more important, and degeneracy consequently lower. These structural
changes contribute to weaken the thermal pulses, and reduce both the duration of
the instability and of the interpulse \citep[for more details,
see][]{siess2010}. 
Our models reproduce these trends, as seen in Table \ref{tab:evol}, and in
Figure \ref{fig:panelevo}.  The most massive EMP stars ($M_\mathrm{ini}\gtrsim
5-6\:$\msun) also experience hot TDU \citep{chieffi2001,herwig2004} during which
the H-burning shell is not completely extinguished and maintains high
luminosities (up to $10^4-10^5$ \lsun) during the thermal pulse. As a
consequence the advance inwards of the base of the convective envelope is
prematurely quenched. The effect of hot TDU can also be seen in Figure
\ref{fig:panelevo}, as $<\lambda>$ values for the $Z=10^{-5}$ models of initial
mass $\gtrsim$ 5 \msun{} tend to be lower than those of higher metallicity of
analogous masses. 

It should be stressed that the TDU efficiency strongly depends on the treatment
of convective boundaries and, specifically, on the implementation of
overshooting (see, for instance, \citealt{freytag1996}).  We note that the TDU
efficiency decreases with increasing M$_\mathrm{ini}$ between $\lambda \sim$ 1
and $\lambda \sim 0.05$. The latter low value corresponds to the 7.5 \msun{}
case computed with Blo95, which has a very short TP-SAGB phase. Some authors
obtained significant TDU for their primordial to very metal-poor massive AGB and
Super-AGB models (\citealt{chieffi2001}, \citealt{herwig2004},
\citealt{lau2008}, \citealt{karakas2010} and \citealt{gilpons2013}), whereas
others (\citealt{gilpons2003,gilpons2005}; \citealt{siess2010} and
\citealt{suda2010}) did not find any TDU. This issue is still controversial,
particularly at the lowest metallicity regime. 

\subsubsection{Hot bottom-burning}

We can also see in Figure \ref{fig:panelevo}  that the maximum temperature of
the base of the convective envelope ($T_\mathrm{BCE}$) tends to increase with
decreasing metallicity. This is once more related to the behaviour of core
masses and, to a lesser extent, to the fact that lower metallicity models have
lower C-abundances in their envelopes and thus need higher temperatures to keep
up the CN-reaction rates required to maintain hydrostatic equilibrium.  In
$Z=10^{-5}$ models with $M_\mathrm{ini} \ge 4$\msun, the temperature at the base
of the convective envelope exceeds $\gtrsim$ 30 MK, and hot bottom burning (HBB) 
sets in \citep[see, e.g.][and references therein]{dellagli2018}. 
The same happens in our  3 \msun{} VW93 model after 10 thermal pulses. Note that
HBB is extremely sensitive to the metallicity, and particularly so in the
stellar mass range considered in this work.

All our models experience an overall increase in $^{12}$C surface abundances,
either during the corrosive SDU or early thermal pulses.  Indeed, despite the
relatively high temperatures at the BCE, \iso{12}C is very efficiently
replenished by TDU during the numerous thermal pulses. In addition, very short
interpulse periods  (Table \ref{tab:evol} and Figure \ref{fig:panelevo})
contribute to the formation of \iso{14}N but does not lead to a significant
destruction of \iso{12}C.

\subsubsection{Effects of mass-loss rates}

Mass-loss rates (which tend to decrease with decreasing metallicity) are the key
factor which determines the main differences between models in Figure
\ref{fig:panelevo}, that is, the variation in duration of the TP-(S)AGB and
interpulse period. 

Unfortunately, the  mass-loss rates is very uncertain for the evolution and
nucleosynthesis of TP-(S)AGB stars, and especially so in the low-Z regime
(\citealt{gilpons2018} and references therein).  EMP stars are more compact and
hotter than higher metallicity stars of similar masses, and thus yield very low
mass-loss rates when using the prescription by VW93, (which depends strongly on
stellar radius and has a negative dependence with the effective temperature). As
shown by \citet{doherty2014b}, using the prescription by Blo95 (with
$\eta=0.02$), which has a very strong dependence on surface luminosity,
dramatically shortens the duration of the TP-(S)AGB phase of Z=10$^{-4}$ stars
with respect to the sequences calculated with VW93.  When the Z=10$^{-5}$ models
are considered with the Blo95 prescription instead of VW93, the duration of the
TP-(S)AGB is shortened by a factor 3 in the 3 \msun{} model, and a factor 5 in
the 7 \msun{} model.
The maximum surface luminosity during the TP-(S)AGB also decreases (see Figure
\ref{fig:panelevo}), because the cores have less time to grow as massive as in
the VW93 case. Because the Blo95 prescription is based on \citet{reimers1975}
formulation, the behaviour of mass-loss rates with luminosity is very similar in
models computed with these prescriptions. Note that the implementation of VW93,
which includes a relatively strong dependence on the effective temperature
yields considerably lower mass-loss rates in the most massive $Z=10^{-5}$ cases,
which are more compact and hotter than their higher Z counterparts. The cooler 7
and 7.5 \msun{} of $Z=10^{-4}$ and $Z= 10^{-3}$ computed with VW93 have average
mass-loss rates similar to the $Z=10^{-5}$ models of the same mass computed with
Blo95. 

Due to the shorter duration of the TP-(S)AGB in models computed with the wind
prescription by Blo95, the number of thermal pulses and the amount of dredged-up
matter during the TDU also decrease (see Table \ref{tab:evol} and Figure
\ref{fig:panelevo}). The effect is naturally more dramatic when the parameter
values $\eta$=0.04 and 1 are used. In these cases the duration of the TP-AGB is
reduced by approximately a factor 7 and a factor 30 respectively for the 3 and 7
\msun{} models.

With increasing mass-loss rate, fewer TDU episodes occur and the surface
enrichment is consequently reduced. As a result, the average efficiency
$\langle\lambda\rangle$ is lower in our sequences calculated with Blo95 compared
to those using the VW93 prescription (see Table \ref{tab:evol} and Figure
\ref{fig:panelevo}). 
We also note that the rise in $<\lambda>$ with thermal pulse number is almost
independent of the mass loss rate prescription.  The maximum efficiency is
reached at about the same pulse number and its value is also very comparable.
Note finally that a shorter TP-(S)AGB phase leads to a lower HBB efficiency,
because it operates over a shorter time and on top of a less massive cores,
which implies lower BCE temperatures. 
We can see in Figure \ref{fig:panelevo}, that the effect of mass-loss rates on
$T_\mathrm{BCE}$ is very mild between 4 and 7 \msun{}, but becomes significant
for our lowest-mass models. The 3 \msun{} models are actually close to the lower
mass threshold required for the occurrence of HBB, and thus, the model
calculated with Blo95, which evolves rapidly on the TP-AGB can not develop an
efficient HBB by the time most of its envelope is lost.

\begin{figure}[ht]
\begin{center}
	\includegraphics[width=1.02\linewidth]{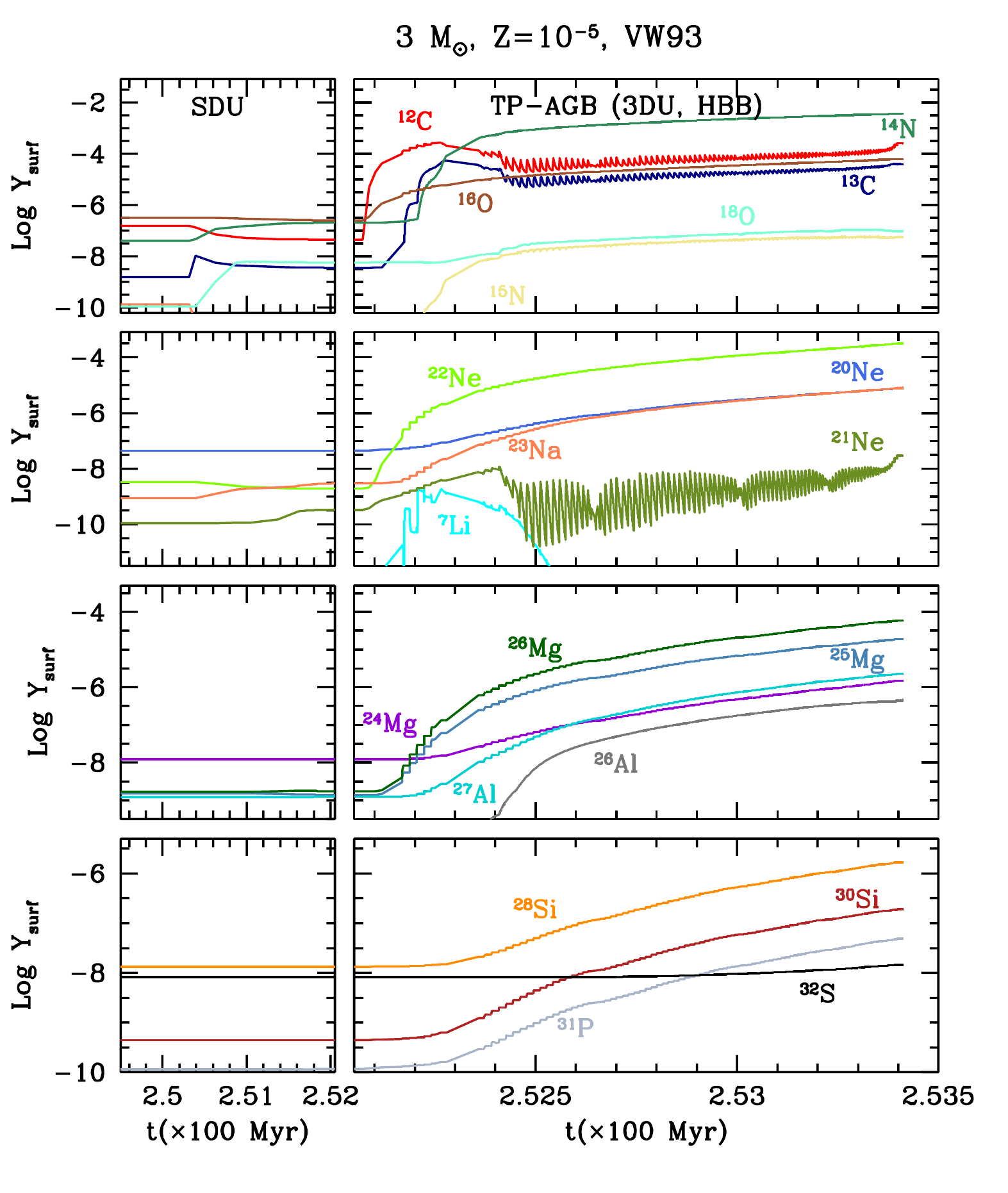}
\caption{Evolution of the surface abundances of some selected isotopes for the
	3~\msun{} $Z=10^{-5}$ model computed with the \citet{vassiliadis1993}
	mass-loss rates.} \label{fig:evoisot3}
\end{center}
\end{figure}

\begin{figure}[ht]
\begin{center}
\includegraphics[width=1.02\linewidth]{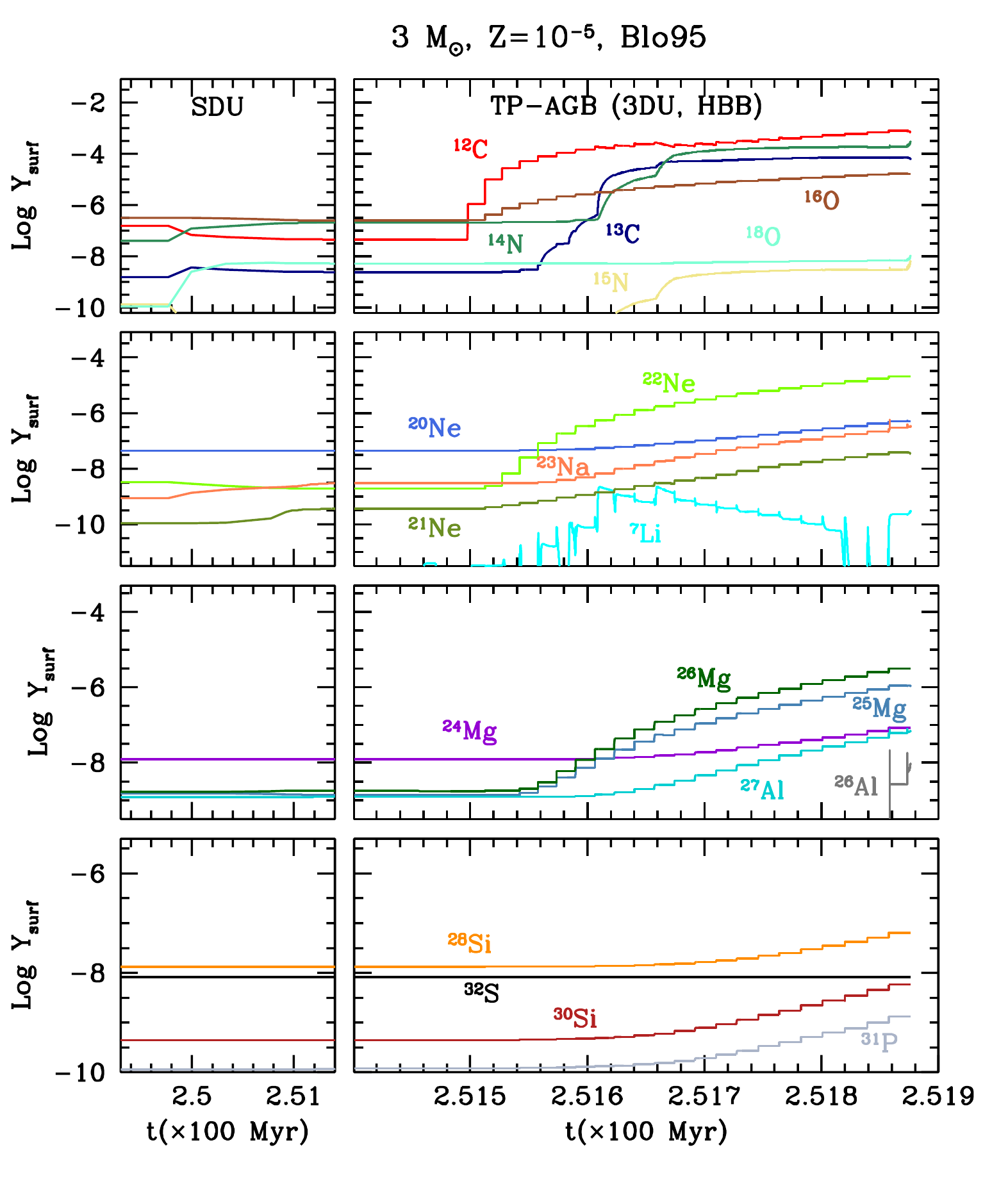} \caption{Evolution
	of the surface abundances of some selected isotopes for the 3~\msun{}
	$Z=10^{-5}$ model, using mass-loss rates as in  \citet{bloecker1995}
	(see main text for details).}
\label{fig:evoisot3B}
\end{center}
\end{figure}

When considering the effects of mass-loss rates on the TP-AGB evolution at
various metallicities (Figure \ref{fig:panelevo}), we must recall that the
mass-loss rates used for the $Z=10^{-4}$ models are an adapted version of
\citet{reimers1975}, which leads to considerably shorter TP-AGB phases than
those resulting from the prescription by VW93. For a comparison, the
$Z=10^{-4}$, 5 \msun{} model calculated with  \citet{reimers1975} undergoes 69
thermal pulses, whereas the same model computed with VW93 undergoes 138 thermal
pulses \citep[see][]{karakas2007}. As a consequence of the  different wind
prescription, the duration of the TP-AGB phase of $Z=10^{-4}$ models becomes
even shorter than that of $Z=0.001$ of analogous masses. 

The upper left panel of Figure \ref{fig:panelevo} shows that the
initial-to-final mass relation of $Z=10^{-5}$ models is less steep than that of
higher metallicity cases.  Due to the significantly longer TP-AGB phases of
$Z=10^{-5}$ cases, the 3 and 4 \msun{}  models develop more massive cores than
their higher $Z$ counterparts, regardless of the wind prescription. We also note
that faster evolving super-AGB models ($M_\mathrm{ini} \gtrsim$ 7 \msun) yield
very similar final core masses that are almost independent of metallicity, when
the same (VW93) prescription is used. The use of Blo95, which leads to even
faster evolution, yields final cores of masses between 0.03 and 0.08 \msun{}
lower.

As a summary, due to the SDU and early TP-(S)AGB evolution, the metallicity (in
terms of Z) is increased to near-solar, and the subsequent TP-(S)AGB evolution
of our initially EMP models is qualitatively very similar to that of metal-rich
stars. However, it is interesting to note some relevant quantitative
differences. Compared to higher metallicity models, EMP stars have: 

\begin{itemize}
\item longer duration of the TP-(S)AGB, determined mainly by the relative
weakness of winds at lower metallicity; 
\item shorter interpulse periods $\Delta t_{IP}$ as a consequence of their
larger core masses; 
\item a higher number of thermal pulses (both because of the shorter $\Delta
t_{IP}$ and the longer duration of the TP-(S)AGB); 
\item thinner intershell mass;
\item higher convective intershell temperatures ($\gtrsim$ 360 MK);
\item higher temperatures at the base of the convective envelope (average
T$_{BCE}$  $\gtrsim$ 33 MK).  
\end{itemize}

We finally note that just like metal-rich stars, model convergence is lost prior
to the complete ejection of the H-rich envelope (remnant envelope mass may be as
high as $\gtrsim$ 1 \msun{} for our more massive models). This failure is
related to the development of an Fe-opacity peak near the base of the convective
envelope, and was analysed in \citet{lau2012}.  It is also interesting to note
that the remaining envelope mass at the end of our calculations is lower when
higher wind rates are used. Indeed, the instability described in
\citealt{lau2012} seems to be favoured by the higher density and temperature at
the BCE which are achieved in models with slower mass-loss rates (and thus more
massive final H-exhausted core). 
The immediate evolution of the star after this instability remains a subject of debate.

\begin{figure}[t]
\begin{center}
\includegraphics[width=1.02\linewidth]{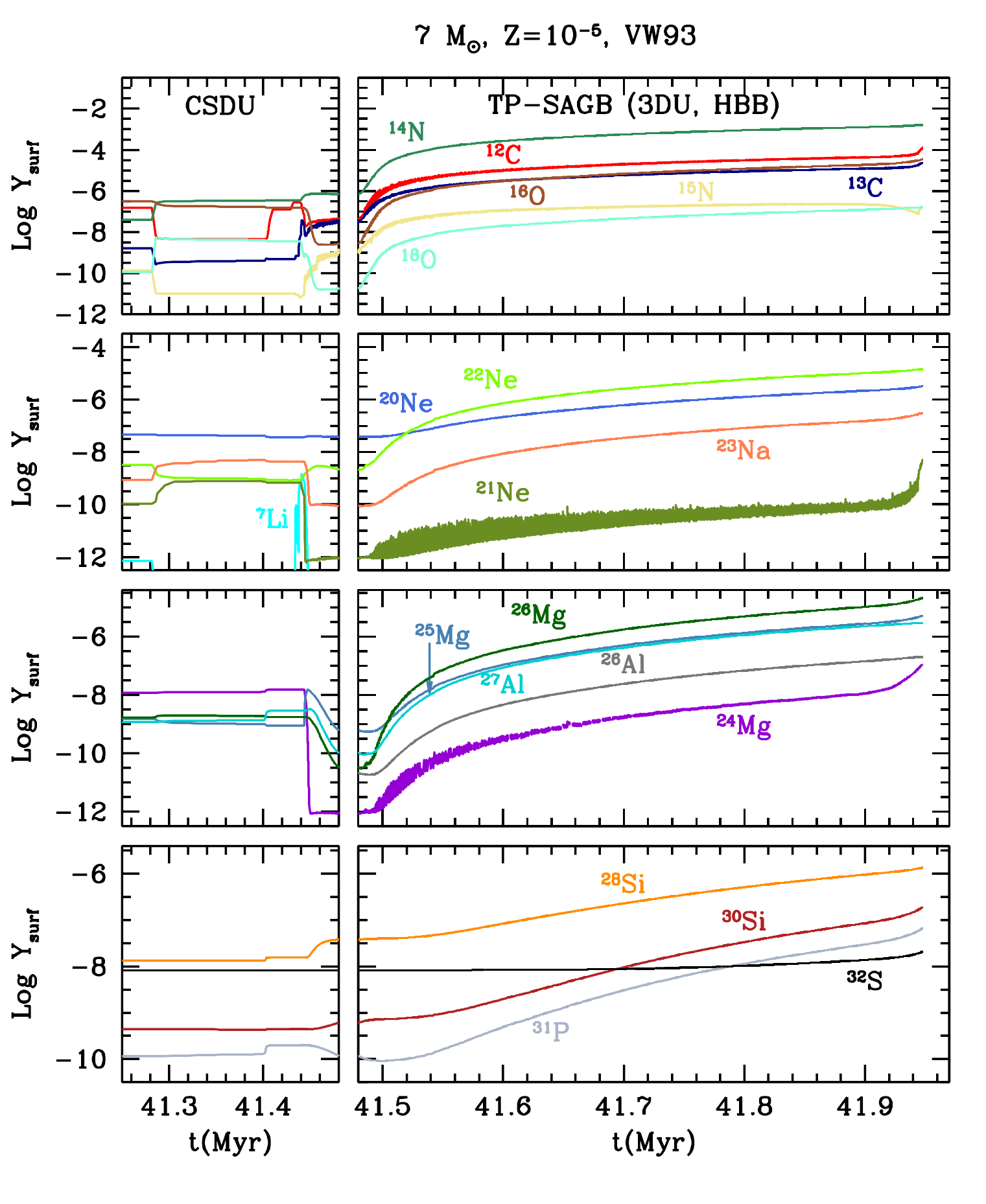}
\caption{Evolution of the surface abundances of some selected isotopes for the
	7~\msun{} $Z=10^{-5}$ model computed with the mass-loss rates by
	\citet{vassiliadis1993}.} \label{fig:evoisot7}
\end{center}
\end{figure}

\begin{figure}[ht]
\begin{center}
\includegraphics[width=1.02\linewidth]{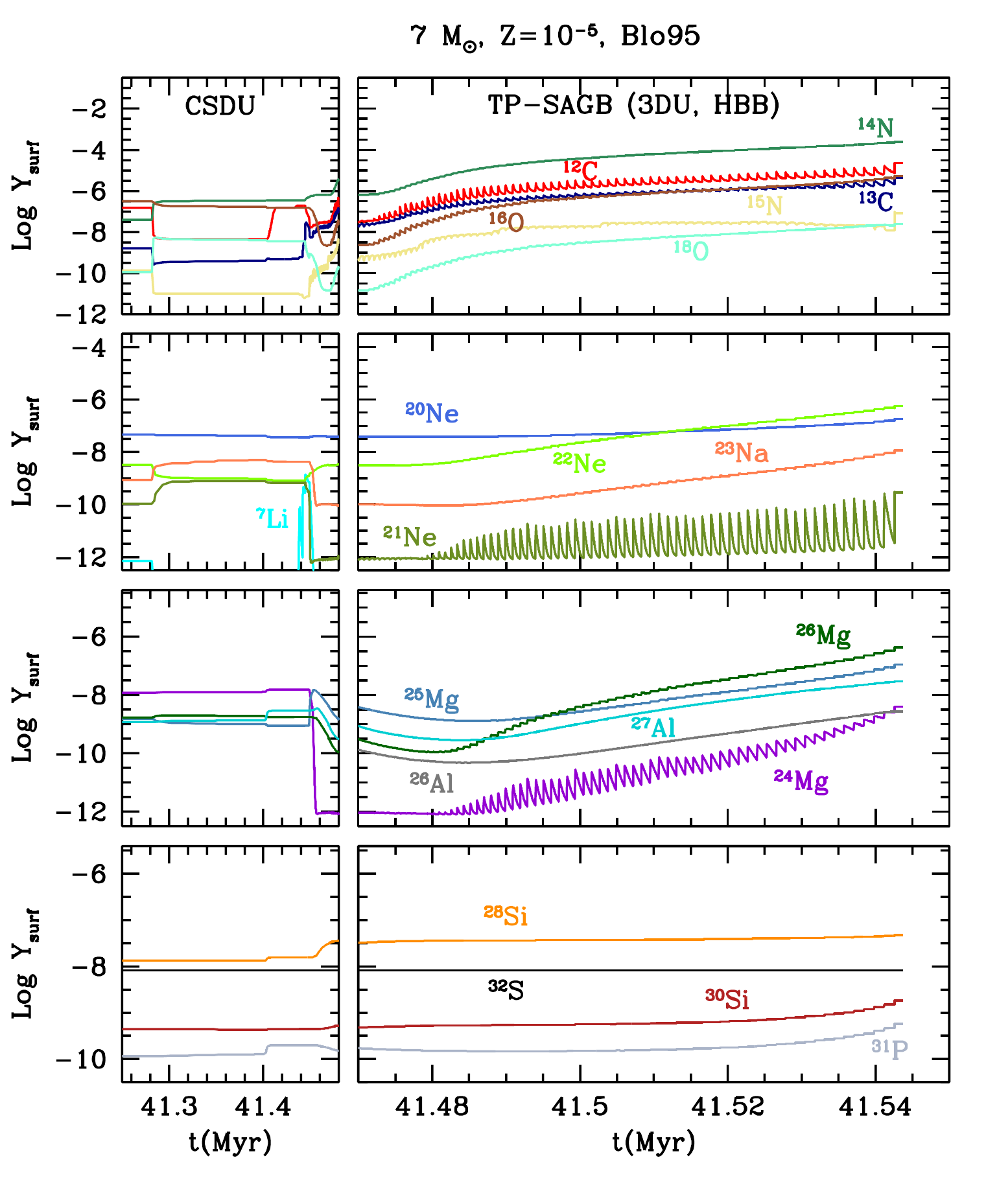} \caption{Evolution
	of the surface abundances of some selected isotopes for the 7.5~\msun{}
	$Z=10^{-5}$ model computed with the mass-loss rates by
	\citet{bloecker1995}.}
\label{fig:evoisot75}
\end{center}
\end{figure}


\section{Nucleosynthesis and evolution of isotopic surface abundances} \label{sec:evonuc}

\subsection{Code description}\label{sec:chemevcode}

Detailed nucleosynthetic calculations were performed using \textsc{monsoon}, the
postprocessing code developed at Monash University
\citep{can93,lug04,doherty2014a}. \textsc{monsoon} takes as inputs the basic
structure profiles of the models computed with \textsc{monstar} (temperature,
density, convective velocity at each mass point). It calculates abundance
variations due to nuclear reaction rates and time-dependant convection using a
‘‘donor cell’’ scheme \citep{can93,henkel17}. \textsc{monsoon} builds its own
mesh-point distribution for each new model, which allows higher resolution in
regions with large abundance variations. 

Nuclear reaction rates are mostly from the \textsc{JINA} reaction library
\citep{cyb10}. p-captures for the NeNa-cycle and MgAl chain are from
\citet{iliadis2001}, p-captures on \iso{22}Ne are from \cite{hale2002},
$\alpha$-captures on \iso{22}Ne are from \citet{karakas2006}, and p-captures on
\iso{23}Na are from \cite{hale2004}. The version of \textsc{monsoon} used for
the present work includes 77 species, up to \iso{32}S and Fe-peak elements.
Additionally, it includes a 'g' particle \citep{lug04}, which is a proxy for
s-process elements. Eventual neutron captures on nuclides which are not present
in our network are accounted for by assigning a cross section to n-captures on
\iso{62}Ni which corresponds to an average of cross sections of n-captures up to
\iso{209}Bi. This neutron-sink approach was used by
\citet{jorissen1989,lugaro2003,herwig2003}.  

Note that the use of a postprocessing code does not allow the feedback of
detailed composition on the evolution.  The effects on the equation of state
(through the mean molecular weight), energy generation and on the opacities are
however expected to be very limited since we are dealing with trace elements.

\subsection{Surface composition changes during the standard and corrosive second dredge-up}

During a standard SDU a very strong surface enrichment in \iso{4}He abundance,
$X_\mathrm{surf}$(\iso{4}He), occurs. Our models have an initial
$X_\mathrm{surf}$(\iso{4}He) = 0.248. In the case of our 3 \msun{} model
computed with VW93, $X_\mathrm{surf}$(\iso{4}He)= 0.277 after the SDU, and 0.308
at the end of the evolution. Surface \iso{4}He enhancement is even higher for
more massive models. For our 6 \msun{} model, $X_\mathrm{surf}$(\iso{4}He)
values at the end of the SDU and at the end of the evolution are, respectively,
0.339 and 0.373.   Because the SDU is so efficient at increasing the surface
\iso{4}He, models computed with Blo95 also produced high yields of this isotope,
in spite of their shorter TP-AGB. This is one of the main reasons why IM stars
can be considered as good candidates for the pollution of the intracluster
medium that gave rise to the formation of multiple stellar populations in
globular clusters, that are characterised by different He over-abundances (see,
e.g. \citealt{milone2012a}, \citealt{milone2014}, \citealt{piotto2012},  and
\citealt{piotto2013}). Note also that helium mass fractions close to 0.4 have
been reported  \citep[e.g.][]{norris2004,piotto2007,bellini2013}.
The effects of SDU on isotopes beyond \iso{4}He are also significant. Regardless
of the wind prescription  $^{14}$N surface abundances at the end of this episode
increase by a factor 5-6 with respect to their initial values.  Surface
abundances of \iso{18}O are moderately enhanced and to a lesser extent
\iso{21}Ne and \iso{23}Na. Simultaneously, \iso{12}C is depleted and the
abundances of \iso{16}O and \iso{22}Ne slightly decreased (Figures ~
\ref{fig:evoisot3} and \ref{fig:evoisot3B}).

Models of initial mass M$_\mathrm{ini} \gtrsim 7$ \msun{} undergo the corrosive
SDU (CSDU). Like a normal SDU, it causes a high increase in He (e.g.
$X_\mathrm{surf}$(\iso{4}He) is 0.364 at the end of the CSDU, and 0.456 at the
end of the evolution of our model computed with VW93). In addition,  mild
enhancements occur for \iso{13}C, \iso{22}Ne (through
\iso{14}N($\alpha,\gamma$)\iso{18}F($\beta^+,\nu$)\iso{18}O($\alpha,\gamma$)\iso{22}Ne),
\iso{23}Na and \iso{27}Al (see Figures \ref{fig:evoisot7} and
\ref{fig:evoisot75}).  Interestingly surface $X_\mathrm{surf}$(\iso{7}Li) is
clearly enhanced. It is created by e-capture on \iso{7}Be which is itself formed
by $\alpha$-capture on \iso{3}He \citep{cameron1971}. Shortly afterwards
\iso{7}Li is destroyed by p-captures. 
The distinctive signature of the corrosive SDU is the additional pollution with
He-burning products, mainly \iso{12}C, which contributes to increase the
envelope opacity and in turn the mass loss rate (see, for instance,
\citealt{marigo2007}, \citealt{nanni2018}). 

\begin{figure*}[ht]
\begin{center}
\vspace{-0.1cm}
\includegraphics[width=0.95\linewidth]{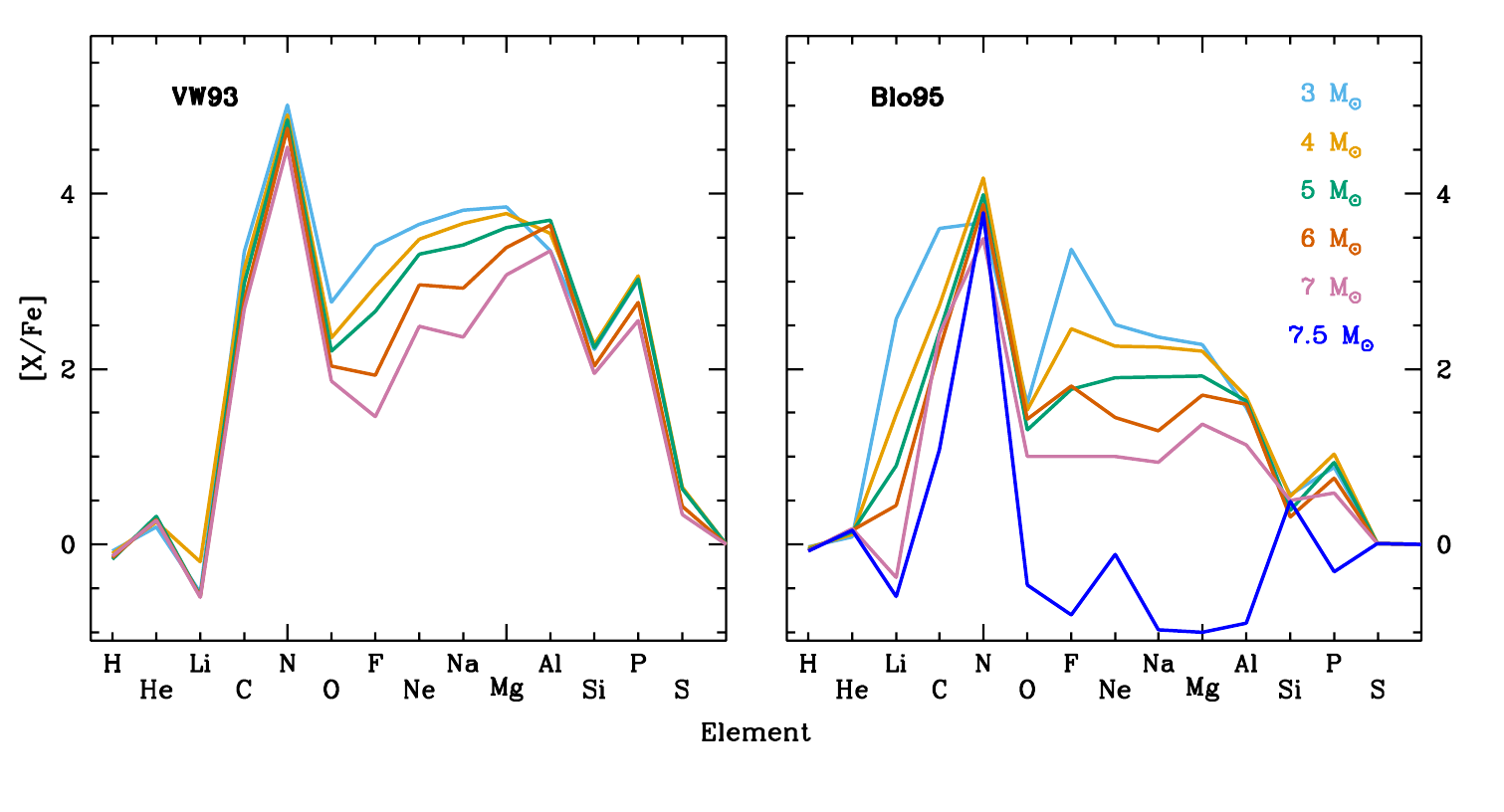}
\caption{Abundance pattern in terms of the ejecta of our models in the [X/Fe]
	notation. Solid lines refer to models computed with VW93, and dashed
	lines refer t those computed with Blo95}. 
	\label{fig:isotopes}
\end{center}
\end{figure*}

\subsection{Nucleosynthesis during the TP-(S)AGB evolution}

\subsubsection{Third dredge-up episode}

The TDU allows the transport to the surface of isotopes synthesised in the
convective zones associated with thermal pulses. All  our models experience TDU
whose efficiency, as shown in Section \ref{sec:evol}, decreases with increasing
initial mass (see Table \ref{tab:evol}).  The TDU raises
$X_\mathrm{surf}$(\iso{12}C) by 3 orders of magnitude for the 3 \msun{} model
calculated with VW93 (which reaches $X_\mathrm{surf}$(\iso{12}C)=$10^{-3}$) and
by 2 orders of magnitude for the 5 \msun{} model calculated with VW93 (up to
$X_\mathrm{surf}$(\iso{12}C)=$2\times 10^{-4}$).

The enhancement in \iso{4}He and \iso{12}C are the most significant signatures
of the occurrence of TDU, but He-burning in the pulse-driven convective zone
also contributes to the synthesis of 
\iso{12}C($\alpha,\gamma$). To a lesser extent, \iso{16}O also forms via
\iso{13}C($\alpha$,n)\iso{16}O and \iso{13}N($\alpha$,p)\iso{16}O, 
and \iso{20}Ne is slightly produced by
\iso{16}O(n,$\gamma$)\iso{17}O($\alpha$,n)\iso{20}Ne.  Besides, \iso{21}Ne is
created through \iso{16}O(n,$\gamma$)\iso{17}O($\alpha,\gamma$)\iso{21}Ne and
\iso{20}Ne(n,$\gamma$)\iso{21}Ne.
\iso{22}Ne forms from
\iso{14}N($\alpha$,$\gamma$)\iso{18}F($\beta^+$,$\nu$)\iso{18}O($\alpha$,$\gamma$)\iso{22}Ne.
\iso{25}Mg and \iso{26}Mg are synthesised via ($\alpha$,n) and ($\alpha,\gamma$)
reactions on \iso{22}Ne, respectively.
It is important to recall that in IM EMP stars the H burning shell remains
active during the development of the thermal pulse, with luminosities up to
$10^4$-$10^5$\lsun. This allows \iso{4}He, \iso{13}C, \iso{14}N and \iso{15}N to
be created above the He-flash driven convective zone while flashes are
occurring. Note however that, because the H-burning and He-burning regions are
not mixed until the subsequent TDU episode occurs, the matter synthesised in the
H burning shell cannot fuel He-burning during the current flash. Neutrons,
relevant for the occurrence of s-process nucleosynthesis, are produced mainly
via \iso{22}Ne($\alpha$,n)\iso{25}Mg \citep{straniero1997} within the thermal
pulse if the temperature at the base of the pulse exceeds $\gtrsim 3.5\times
10^8$K.  This source is activated in our most massive Z$=10^{-5}$ models.  Right
panels of Figures \ref{fig:evoisot3} to \ref{fig:evoisot75}  show the evolution
abundances of selected isotopes during the TP-(S)AGB phase.

\begin{figure*}[ht]
\begin{center}
\includegraphics[width=0.8\linewidth]{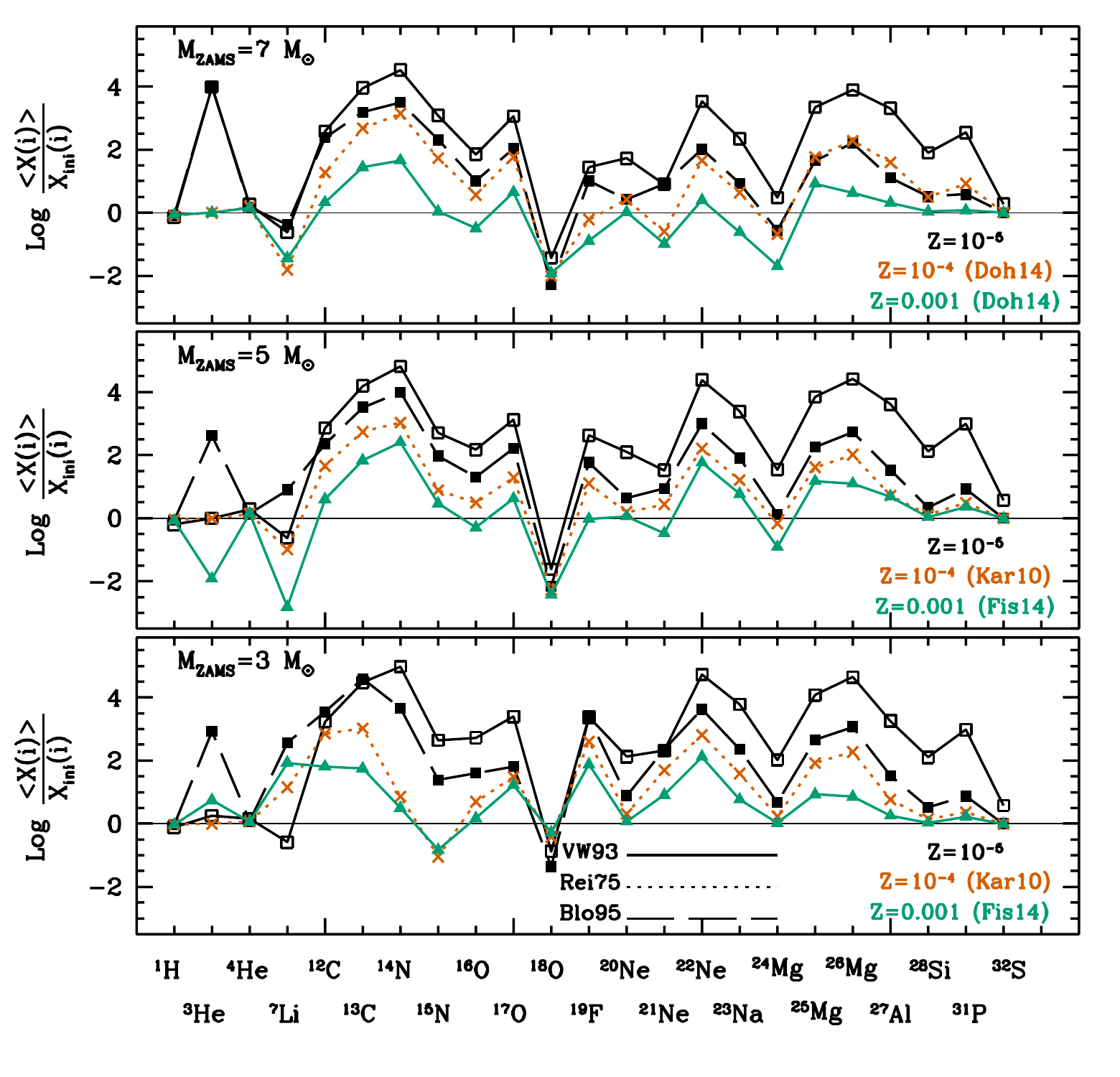}
\caption{Production factor of selected isotopes for the low metallicity models
	of 7, 5 and 3 \msun. The $Z=10^{-5}$ results (black) are compared with
	the $Z=10^{-4}$ calculations (orange) of \citet{karakas2010} for the
	3\msun{} and 5 \msun{} cases, and \citet{doherty2014b} for the 7\msun.
	Results are also compared with the Z=0.001, 3 and 5 \msun{} models by
	\citet{fishlock2014}, with and the 7 \msun{} model by
	\citet{doherty2014b}. All models were calculated with the same
	\citet{vassiliadis1993} wind prescription (VW93), except for the ones
	shown with dotted and dashed lines, which were calculated with
	\citet{reimers1975} (Rei75) and \citet{bloecker1995} with $\eta$=0.02
	(Blo95).  }
\label{fig:compareprod}
\end{center}
\end{figure*}

\subsubsection{Hot bottom burning}

The temperature at the base of the convective envelopes of our model stars is
between 30 MK and $\gtrsim$ 140 MK, and thus their TP-AGB phase is mostly
dominated by the occurrence of the HBB, in particular by the onset of the CN and
NeNa-cycles and the MgAl-chain.  Note however that, as mentioned above, our
lowest M$_\mathrm{ini}$ case (3 \msun{}) does not develop significant HBB until
its envelope has been sufficiently enriched in metals ($Z_{env} \lesssim
10^{-3}$) due to efficient early TDU episodes (see Figures \ref{fig:evoisot3}
and \ref{fig:evoisot3B}).

From the nucleosynthetic point of view, the main result of HBB is an enrichment
of the stellar envelope in \iso{4}He (which was also enhanced by SDU). The onset
of the CN-cycle also produces a significant increase in \iso{14}N and \iso{13}C,
and milder enhancements in \iso{15}N, at the expense of \iso{12}C, \iso{16}O and
\iso{18}O.  Despite the occurrence of HBB, overall the surface abundances of
\iso{12}C and \iso{16}O reach near equilibrium values and even increase along
the TP-(S)AGB phase because of the efficient TDU replenishing these isotopes.
Also, the surface \iso{12}C/\iso{13}C ratio remains nearly constant around its
equilibrium value of 4 during most of the TP-(S)AGB evolution, and all our stars
become C-rich.  

Through the NeNa-cycle the surface abundance of \iso{20}Ne increases at the
expense of \iso{21}Ne, \iso{22}Ne, and \iso{23}Na abundances.  Because $T_{BCE}$
values are higher (and thus the NeNa-cycle more efficient) for more massive
models, \iso{20}Ne enhancement is more significant. This can be seen by
comparing Figures \ref{fig:evoisot3} and \ref{fig:evoisot7}. 
\iso{21}Ne is quickly destroyed by p-capture during HBB, regardless of the
initial mass. \iso{22}Ne is converted into \iso{20}Ne
(\iso{22}Ne(p,$\gamma$)\iso{23}Na(p,$\alpha$)\iso{20}Ne), and into \iso{26}Mg by
$\alpha$-captures. However, \iso{22}Ne is efficiently replenished by TDU raising
its surface abundance and feeding the NeNa-cycle.  The implementation of the
fastest Blo95 mass-loss rate decreases the time during which the NeNa-cycle is
active. It also reduces the number of TDU episodes, the final core mass and
incidentally $T_{BCE}$. As a consequence, the enhancement of \iso{23}Na  all the
Ne isotopes is significantly diminished. This can be seen by comparing Figures
\ref{fig:evoisot3} and \ref{fig:evoisot3B}, which show the surface abundance
evolution of 3 \msun{} models with VW93 and Blo95 respectively, and by comparing
\ref{fig:evoisot7} and \ref{fig:evoisot75}), which show analogous information
for the 7 \msun{} cases.

The temperatures at the BCE are also high enough for the activation of the
MgAlSi-chain. \iso{26}Al and \iso{27}Al increase at the expense of Mg isotopes
via subsequent proton captures and $\beta$-decays.
\iso{24}Mg(p,$\gamma$)\iso{25}Al($\beta^+$,$\nu$)\iso{25}Mg(p,$\gamma$)\iso{26}Al(p,$\gamma$)\iso{27}Si($\beta^+$,$\nu$)\iso{27}Al.
All our models computed with VW93 reach $T_{BCE}$  high enough to allow for the
formation of significant amounts of Al and Si isotopes, and the higher the
initial mass (and thus the average $T_{BCE}$), the higher the Al and Si yields.
Note that these isotopes are relevant for the formation of grains and can
ultimately the impact stellar wind.  As it happens with the NeNa-cycle, the
shorter TP-(S)AGB duration and lower $T_{BCE}$ of Blo95 models significantly
reduce the yields of all the isotopes involved in the MgAlSi-chain.
It is important to recall that, given the high $T_{BCE}$ values of our models,
uncertainties in the rates of the reactions involved in the NeNa-cycle and in
the MgAlSi-chain are expected to have an important effect on the corresponding
nucleosynthetic yields \citep{izzard2006}, specially on those of \iso{22}Ne,
\iso{23}Na and \iso{26}Al.

With higher mass loss rate prescriptions, the effect of HBB is reduced mainly
because of the shorter duration of the TP-(S)AGB phase. The enrichment in
\iso{14}N and, to a less extent, of \iso{13}C and \iso{18}O are smaller.
Actually, both TDU and HBB acting during shorter times, significantly reduce the
surface enhancement of all isotopes above \iso{20}Ne, with respect to the cases
computed with VW93. Note also that \iso{12}C is only weakly affected by the
different wind prescriptions because its abundance is maintained at its
equilibrium value.  Figures \ref{fig:evoisot3} and \ref{fig:evoisot3B} show the
effect of different mass-loss rates on the surface abundance evolution of the
most abundant isotopes. 


\section{Nucleosynthetic yields}\label{sec:yields}

Net nucleosynthetic yields of isotope $i$ are expressed in \msun{} and
calculated as follows: 
\begin{equation}
    M_i=\int_0^{t_\mathrm{end}}{\left[ X_i(t)-X_{i, ini}\right ]\dot M(t)dt}
    \label{eq:yields}
\end{equation}
\noindent where $t_\mathrm{end}$ is the time at the end of our calculations,
$X_{i, ini}$ is the initial mass fraction of isotope $i$, $X_i(t)$ its value at
an arbitrary time $t$ and $\dot M(t)$ is the mass-loss rate at $t$.

As mentioned in Section \ref{sec:evotp}, due to the occurrence of an
opacity-related instability, some envelope mass is left at the end of our
calculations. The further evolution of stars at this point is unknown, and they
might either recover stability and proceed along some more thermal pulses, or
quickly eject their remaining envelopes. We assume the latter happens, and the
corresponding contribution to the yields is calculated by assuming no further
nuclear processing, that is, by substracting the mass of each isotope in the
remaining envelope minus the mass of that isotope which a remaining envelope of
the same mass and with initial composition would have.

The expected abundances associated to the ejected matter are presented in Table
\ref{tab:abuns} and in Figure \ref{fig:isotopes}.  The most striking feature in
this figure points to the key role of the highly uncertain mass-loss rates in
this metallicity regime. As described in Section \ref{sec:evonuc}, all our
models produce significant amounts of He, C and N due to the combination of HBB
and TDU. Models computed with the VW93 mass-loss treatment experience longer
TP-(S)AGB phases and longer (and more efficient) CNO, NeNa and MgAl processing
as well as more dredge-up  episodes. As a consequence, their abundance patterns
(in terms of their ejecta) for elements between Ne and P are 1 to 2 orders of
magnitude higher than those of their analogous cases calculated with Blo95. Note
in addition the trend with initial masses. In general, lower mass models (which
also experience longer TP-(S)AGB phases) yield higher abundances.

Detailed yield tables, together with mass lost, initial and average abundances,
and production factors for each of our 77 species are presented in the
Supplementary Information in the same format as \cite{doherty2014a,doherty2014b}
and available on-line \url{http://dfa.upc.es/personals/pilar/research.php}.

\begin{figure*}[ht]
\begin{center}
\includegraphics[]{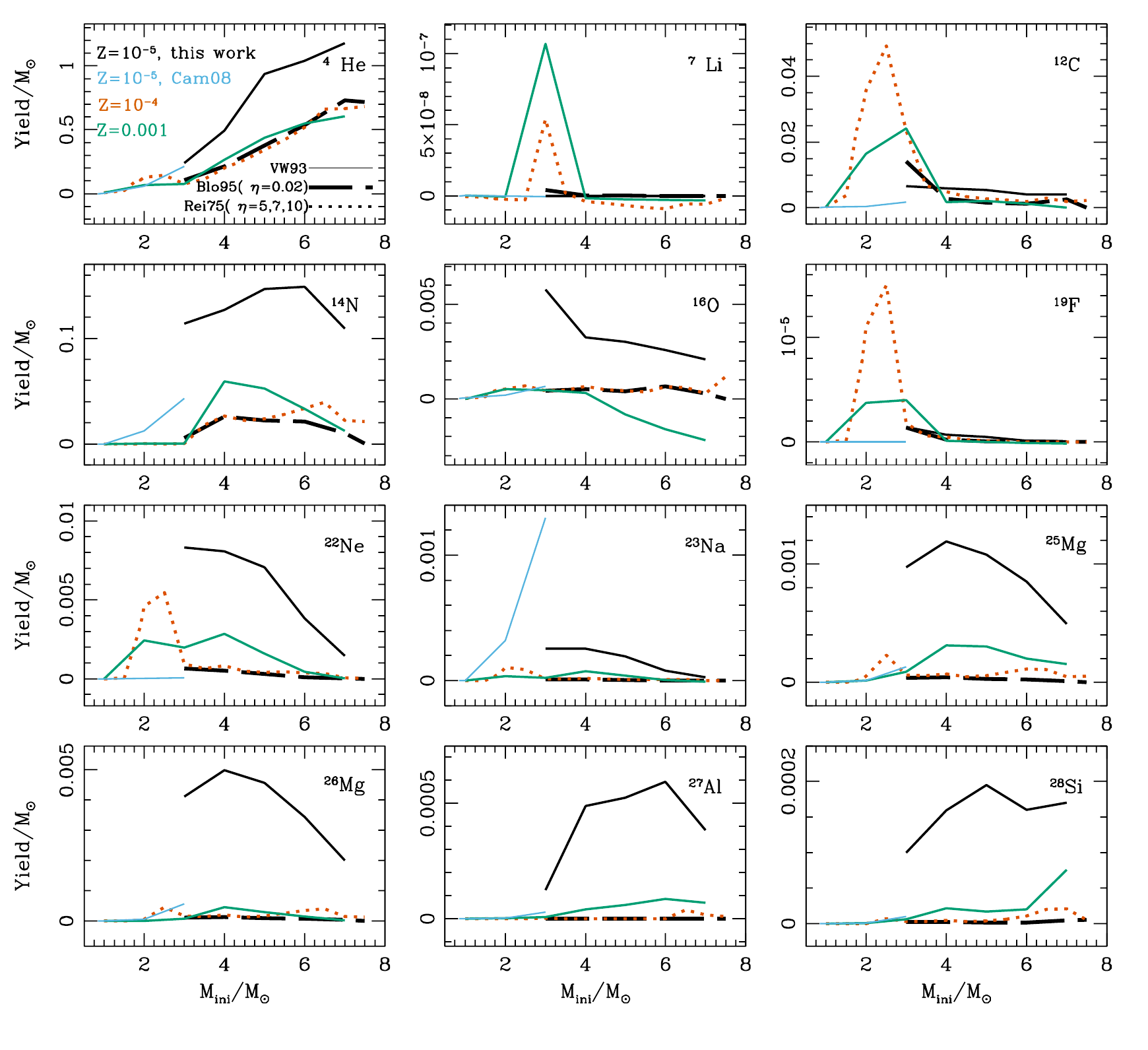}
\caption{ Yields of intermediate-mass stars of different metallicities.  Values
	for $Z=10^{-5}$ cases are shown in thin blue line for the models
	calculated with VW93 from \citet{campbell2008} (0.85, 1, 2 and 3 \msun).
	Our models computed with VW93 (between 3 and 7 \msun) are shown in black
	solid lines, and those with Blo95 are shown in black dashed lines.
	Values in orange correspond to the $Z=10^{-4}$ models from
	\citet{karakas2010} (1 to 6 \msun) and \citet{doherty2014b} (7 and 7.5
	\msun). Values in green correspond to the $Z=10^{-3}$ models from
	\citet{fishlock2014} (1 to 6 \msun) and \citet{doherty2014b} (7 and 7.5
	\msun).}
	\label{fig:panyields}
\end{center}
\end{figure*}

The production factor for each species $i$, is defined as 
\begin{equation}
    P_i=\log{\frac{\langle X_i\rangle}{X_{i,\mathrm{ini}}}}
\end{equation}
where $\langle X_i\rangle = \frac{1}{t_{{\rm star}}}\sum_{j=1}^N X_{ij}(t)
\Delta t_j$ with $t_\mathrm{star}= \sum  \Delta t_j$ is the age of the star,
$\Delta t_j$ is the duration of time step $j$ and $X_{ij}(t)$ the surface
abundance of isotope $i$ at that time.   Figure \ref{fig:compareprod} compares
production factors for EMPs and very metal-poor models of different initial
masses. We included the results from our work, as well as the $Z=10^{-4}$ models
of \citet{karakas2010} and \citet{doherty2014b}, and the $Z=0.001$ models by
\citet{fishlock2014} and \citet{doherty2014b}. 
The overproduction factors show the effects of the TDU and HBB nucleosynthesis
described in detail in Section \ref{sec:evonuc}. Specifically, they highlight
the increasing efficiency of TDU and HBB for decreasing metallicity.  For
instance, the production factors of \iso{15}N, \iso{17}O and \iso{21}Ne are
negative for the  $Z=10^{-3}$ 5 and 7 \msun{} models, and positive otherwise. 
The reason is that, in metal poor stars, the initial isotopic abundances are
lower and the temperature of nuclearly active layers hotter, the dredge-up
efficiency higher and the duration of their TP-(S)AGB phase longer. 

As explained in Section \ref{sec:evonuc}, with a higher mass loss rate (Blo95)
the TP-(S)AGB is considerably shortened and the TDU and HBB efficiencies lower,
so the production factors are generally reduced with respect to the calculations
by VW93. However, \iso{12}C yields are practically the same with Blo95, as the
abundance of this isotope is maintained at its equilibrium value.  \iso{3}He and
\iso{7}Li production factors also increase, both for the 3 and the 5  \msun{}
models calculated with Blo95. \iso{7}Li is produced during HBB through the
Cameron-Fowler mechanism \citep{cameron1971}, but easily destroyed (through
\iso{7}Li(p,$\alpha$)$\alpha$) at temperatures as low as 3.5 MK. 
Models computed with Blo95 destroy \iso{7}Li less efficiently because of the
relative shortness of their evolution. 

Lithium abundance analysis is considered an important diagnosis tool for the
understanding of globular cluster chemical enrichment processes
\citep{dorazi2014} although, as pointed out by \citealt{ventura2010}, from the
theoretical point of view, its nucleosynthetic yields are strongly affected by
input physics uncertainties.  Specifically, the behaviour of \iso{7}Li could be
relevant for the understanding of multiple-population low-mass globular
clusters, as many of them show Li production and, furthermore, have first and
second generation stars which share very similar Li abundances (see, e.g. M12 in
\citealt{dorazi2014}, or NGC 362 in \citealt{dorazi2015}, and references
therein). 

For the sake of comparison, let us recall that the FRUITY database
\citep{cristallo2015} also includes the detailed nucleosynthesis yields of
intermediate-mass models of Z=10$^{-4}$. These authors used the wind rates
proposed by \citet{straniero2006} and introduced an $\alpha$ enhancement which
was not present in the calculations by \citet{karakas2010}, and directly affect
the yields of these elements. In addition, HBB in the FRUITY models is less
efficient than in \citet{karakas2010}.
\citet{ritter2018} compared NuGrid model yields with those from
\citet{herwig2004}, \citet{karakas2010} and FRUITY. These authors pointed out to
a coincidence of results within a 2-3 factor, and  justified yield differences
on the bases of convective boundary mixing prescriptions and resolution
affecting the treatment of HBB.


\section{Preliminary exploration of the implications on Galactic chemical evolution}\label{sec:chemev}

As mentioned in the Introduction, there are no detailed nucleosynthetic yields
of intermediate-mass stars of $Z\lesssim10^{-5}$ in the literature apart from
the work by \citet{iwamoto2009}.  Given the uncertainties in the input physics
in this metallicity regime, and its  critical effects on the yields, the
possible effects of $Z=10^{-5}$ intermediate-mass stars on GCE are poorly known.
In this section we compile our results and those obtained with similar versions
of the codes described in Sections \ref{sec:evol}  and \ref{sec:evonuc}.
Specifically, in Figure  \ref{fig:panyields} we present our yields for the
isotopes with the highest production factor values (see Figure
\ref{fig:compareprod}), and compare these to the $Z=10^{-4}$ cases from
\citet{karakas2010} and \citet{doherty2014b}, and the $Z=0.001$ models by
\citet{fishlock2014} and \citet{doherty2014b}. We also include the yields from
the $Z=10^{-5}$, 0.85, 1, 2 and 3 \msun{}  models by \citet{campbell2008}.  

This figure further illustrates the strong effect of the mass-loss rates
prescriptions (recall that all models use the prescription by VW93, except the
additional set of $Z=10^{-5}$ models computed with Blo95, and the $Z=10^{-4}$
models, which use \citet{reimers1975}).  As different wind prescriptions cause
higher variations in yields than metallicity effects, it is difficult to
determine clear trends with metallicity.
Compared to higher metallicity cases, $Z=10^{-5}$ models computed with VW93
globally produce significantly higher yields of \iso{4}He, \iso{14}N, \iso{16}O,
\iso{22}Ne, the heavy Mg isotopes, \iso{27}Al and \iso{28}Si. Specifically, such
high \iso{14}N production, which is already expected in AGB stars of
[Fe/H]$\gtrsim$-2.5 (\citealt{vincenzo2016}, \citealt{kobayashi2020}), may help
explain observations of EMP stars (see, e.g. \citealt{spite2006}).  The
production of \iso{27}Al and \iso{28}Si may impact the formation of dust that
drive the winds of intermediate-mass EMP stars.
As a reference, our $Z=10^{-5}$ models (computed with the prescription by VW93)
dredge-up a total amount of matter between 5 and 8 times higher than the
$Z=10^{-4}$ cases by \citet{karakas2010}, which were computed with the fast
winds provided by \citet{reimers1975} with $\eta=5,7,10$.  Yields of isotopes
which tend to reach equilibrium values (such as \iso{12}C), or which are very
fragile, such as \iso{7}Li and \iso{19}F, increase in the mass range 2-4 \msun{}
when efficient winds are considered. 
One could expect that, if the $Z=10^{-4}$ models had been computed with the less
efficient prescription by VW93, the associated yields would be between the
$Z=10^{-5}$ and the $Z=0.001$ cases displayed. We also note that the great
similarly between the  $Z=10^{-4}$ yields of \citet{karakas2010} and the
$Z=10^{-5}$ ones using Blo95.

Understanding the evolution of the yields at $Z=10^{-5}$ between the  0.85 and 3
\msun{} models of \citet{campbell2008}, and those computed here is nearly
impossible given the significant differences in input physics.
\citet{campbell2008} used the Schwarzschild criterion for convective boundaries,
which yields inefficient or no TDU at all and also used constant composition
low-temperature opacities, which strongly impact the mass loss rate
\citep{marigo2002}. In addition, nuclear reaction rates during the
post-processing calculations were different. \citet{campbell2008} considered the
\textsc{reaclib} data, based on a 1991 update of the compilation by
\citet{thielemann1986} (see \citealt{lug04}), whereas we used the new rates
described in Section \ref{sec:chemevcode}. This is expected to affect
significantly yields of isotopes involved in the NeNa-cycle and the MgAl-chain
and, specially, \iso{23}Na yields  \citep{karakas2010}. These differences in the
input physics introduce strong discontinuities in the yield of a given element
as a function of initial mass as can be seen in Figure ~\ref{fig:panyields}.

The actual contribution of our model stars to Galactic chemical evolution would
depend on the IMF which, unfortunately, is poorly known at the EMP metallicty
regime. Note that the primitive IMF might not be as strongly biased towards low
masses as the present-day IMF (see e.g. \citealt{salpeter1955},
\citealt{scalo1979}, or \citealt{kroupa2001}).    \citealt{suda2013} suggested
that the Galactic IMF might have undergone a transition from massive star to
low-mass star dominated at [Fe/H]$\sim -2$. If this were the case, the
contribution to the Galactic chemical inventory of our model stars (above
3\msun) might be higher than expected from standard IMFs.

\begin{figure}[ht]
\begin{center}
\vspace{-0.2cm}
\includegraphics[width=.99\linewidth]{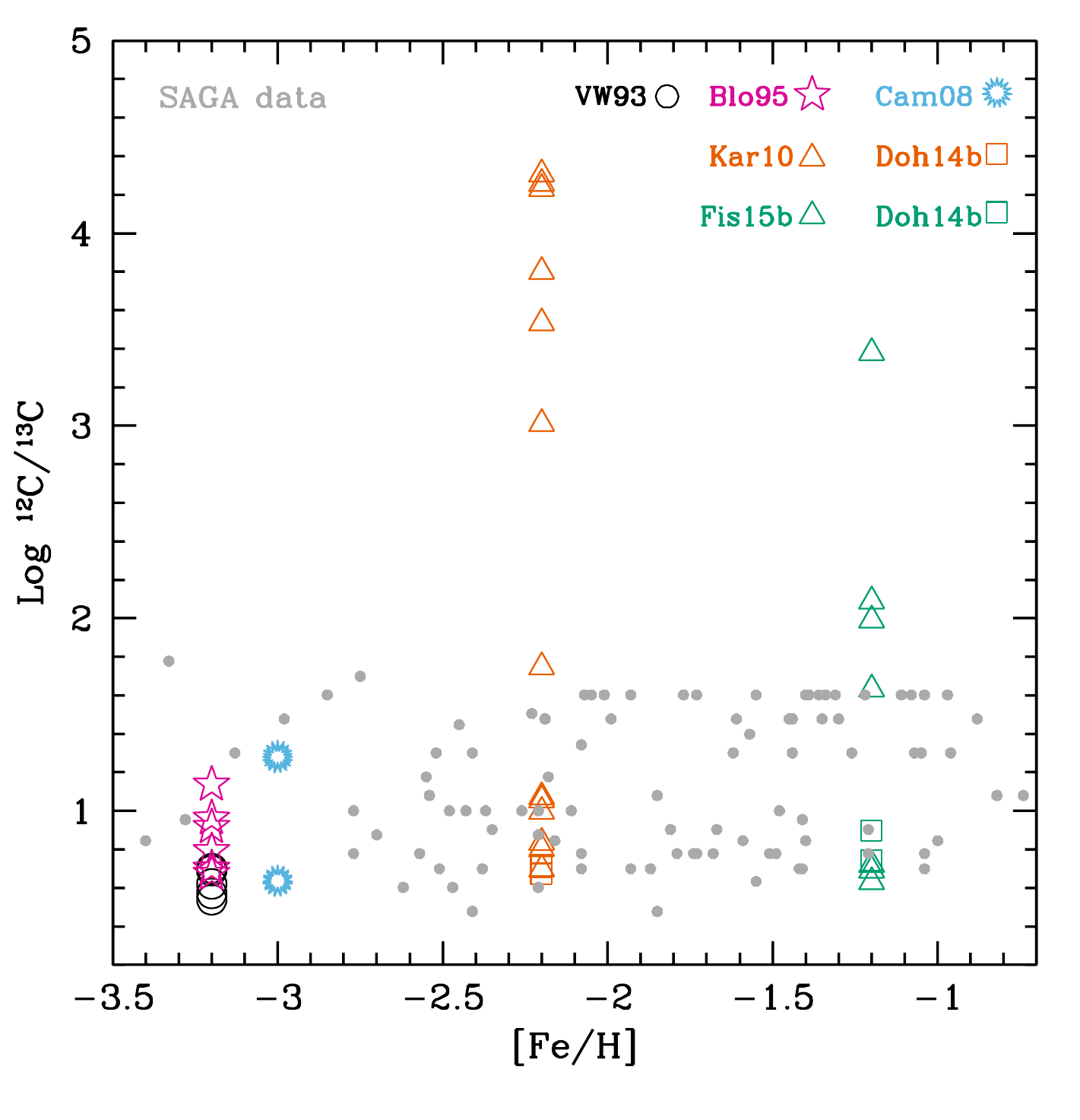}
\caption{Averaged carbon isotopic ratios versus metallicity resulting from our
	model stars using VW mass-loss prescription (black circles), and Blo95
	with $\eta=0.02$ (pink stars); from \citet{campbell2008} (blue stars);
	from \citet{karakas2010} (orange triangles) and \citet{doherty2014b}
	(orange squares); from \citet{fishlock2014} (green triangles) and from
	\citet{doherty2014b} (green squares). Grey symbols represent the
	observed ratios obtained from the SAGA database \citep{suda2008}. }
	\label{fig:isorat}
\end{center}
\end{figure}

When looking at the Li abundances (Table \ref{tab:abuns}) we can see that most
lie in the range of observed values ($0 \lesssim \log\: \epsilon_{{\rm obs}}
(\mathrm{Li}) \lesssim 2.8$), and only our 3 \msun{} case computed with Blo95
lies above the Spite plateau ($\log\: \epsilon (\mathrm{Li}) = 2.2$). In
general, lowest mass models yield higher $\log\: \epsilon (\mathrm{Li})$,
because the temperatures at the base of the convective envelope of the former
are lower.  Note however that the mass-loss rate prescription is, once more,
critical for the $Z=10^{-5}$ models, as those computed with Blo95 allow the
removal of most of the convective envelope before \iso{7}Li abundances
substantially decrease  
(see \citealt{ventura2010}) for a fuller discussion.

Comparison between theoretical results and observationally determined isotopic
ratios has proved an important tool for Galactic Chemical evolution, as
different isotopes of the same element may be formed through different processes
or in stars of different initial masses (see, for instance,
\citealt{kobayashi2011}, \citealt{romano2017}).  When considering
intermediate-mass model results, isotopic ratios of \iso{12}C/\iso{13}C are
useful to determine the relative importance of TDU (which efficiently transports
\iso{12}C), and HBB (which destroys \iso{12}C and produces \iso{13}C at the base
of the convective envelope in AGB stars of initial mass $\gtrsim$ 3-4\msun{}).
\iso{12}C is also significantly produced in massive stars \citep{kobayashi2011},
and \iso{13}C might also be produced in fast-rotating massive stars (see, e.g.
\citealt{meynet2002},  \citealt{chiappini2008}). In order to illustrate how
massive AGB and Super-AGB stars contribute to this isotopic ratio as a function
of metallicity, we present in Figure \ref{fig:isorat} the resulting ratio
patterns in terms of the ejecta of \iso{12}C and \iso{13}C, for metal-poor
models between $Z=10^{-5}$ and $Z=0.001$ and compare them with data from the
SAGA database \citep{suda2008}. Unfortunately, the number of observations for
[Fe/H]$\lesssim$ -2.5 is small. We note that the [Fe/H]=-3.2 and [Fe/H]=-3
models cover reasonably well the range of observations (except, perhaps, for the
highest mass and the VW93 cases), and account for the observed spread in
\iso{12}C/\iso{13}C. 

 Magnesium is another of the few elements for which isotopic ratios may be
 determined (see, e.g. \citealt{gay2000}, \citealt{yong2003},
 \citealt{melendez2009}, \citealt{agafonova2011}). Comparisons between GCE
 models and observations have been used, for instance, to constrain star
 formation rates in cosmological timescales \citep{vangioni2019}, to determine
 the onset of the contribution of AGB stars to GCE
 \citep{fenner2003,melendez2007}, to establish the timescale for the formation
 of the Galactic Halo \citep{Carlos2018}, or to understand self-enrichment in
 Globular Clusters \citep{ventura2018}. All three stable magnesium isotopes can
 be formed in both massive and intermediate-mass stars, however massive stars
 are not able to produce the heavier isotopes \iso{25,26}Mg in significant
 amounts \citep[see][]{timmes1995, alibes2001}, as opposed to low-metallicity
 AGB stars \citep{karakas2003}.
\citet{fenner2003} concluded that GCE models need to include metallicity
dependant AGB yields in order to reproduce Mg isotopic ratios observations for
[Fe/H] $<$ -1. \citet{melendez2007} added additional metal-poor stars to the
sample of Mg isotopic ratio observations and compared to GCE models to establish
that AGB stars did not contribute to the Galactic halo until metallicities
[Fe/H] $\gtrsim$ -1.5.  \citet{vangioni2019} recently confirmed that AGB stars
are the main contributors to the heavy Mg isotopes, and that the agreement
between models and observations down to [Fe/H]=-2.5 improved when
intermediate-mass stars were considered, specially if their mass range was
restricted towards masses between 5 and 8 \msun{}. 

As expected, our $Z=10^{-5}$ models produce massive amounts of the heavy Mg
isotopes compared to \iso{24}Mg, specially when the less efficient wind rates by
VW93 are used. For instance, the relation \iso{24}Mg:\iso{25}Mg:\iso{26}Mg is
1:27:113 for our 5 \msun{} model computed with VW93, and 1:18:60 for the model
of the same initial mass computed with Blo95.  The 7.5 \msun{} model computed
with Blo95, which has a very short TP-SAGB phase, yields a Mg isotopic relation
1:12:10.  The $Z=10^{-4}$ 5 \msun{} model by \citet{karakas2010} also has Mg
isotopic ratios favouring the heavy isotopes: 1:8:23. As a reference, observed
\iso{25,26}Mg/\iso{24}Mg ratios tend to decrease with metallicity, 
which attests of an early production of \iso{24}Mg by massive stars. Note
however that there are very few determinations of Mg isotopic ratios below
[Fe/H] =-2, and no observation at all below [Fe/H]=-2.5.  We expect that future
campaigns will go beyond this threshold. Comparison between GCE models and
observations of Mg isotopic ratios might be used to further constrain input
physics, such as the elusive stellar wind rates in EMP IM stars, just as they
helped to constrain rotation in massive models \citep{vangioni2019}.


\section{A binary scenario for the formation of EMP stars} \label{sec:binary}

To assess the possibility that our model stars were the evolved companions of
currently observed CEMP stars, we revisited the analysis developed in
\citet{gilpons2013}, based on the binary formation scenario proposed by
\citet{suda2004}. Observed CEMP stars have low masses ($\lesssim$ 1 \msun), and
provided the initial orbital period is long enough, the low-mass secondary will
be able to accrete a fraction of the wind material ejected during the TP-(S)AGB
phase of the companion and avoid common envelope evolution.

\begin{figure}[ht]
\begin{center}
\includegraphics[width=1.\linewidth]{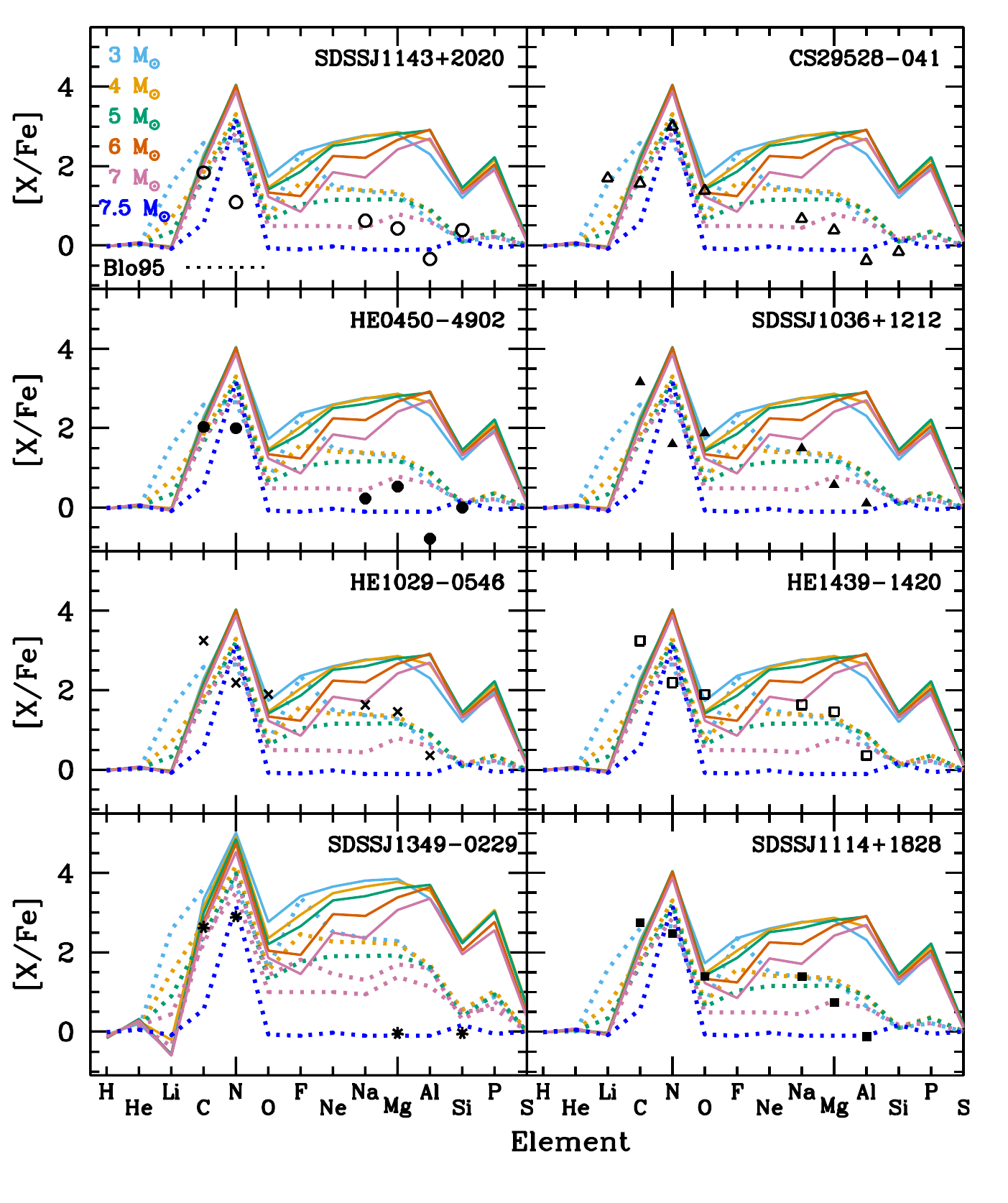}
\caption{Abundance pattern in terms of the ejecta of selected elements. These
	values correspond to a hypothetical case in which 1$\%$ of the matter
	ejected by our evolved models were homogeneously mixed in the 0.2 \msun
	envelope of an unevolved low-mass $Z=10^{-5}$ star (see main text for
	justification). Black symbols correspond to observed abundances of
	selected CEMP-s stars from the SAGA database \citet{suda2017a}. Models
	computed with \citet{vassiliadis1993} and with \citet{bloecker1995} are
	shown, respectively, in solid and dotted lines. } \label{fig:panel_obs}
\end{center}
\end{figure}

\begin{figure}[ht]
\begin{center}
\includegraphics[width=.96\linewidth]{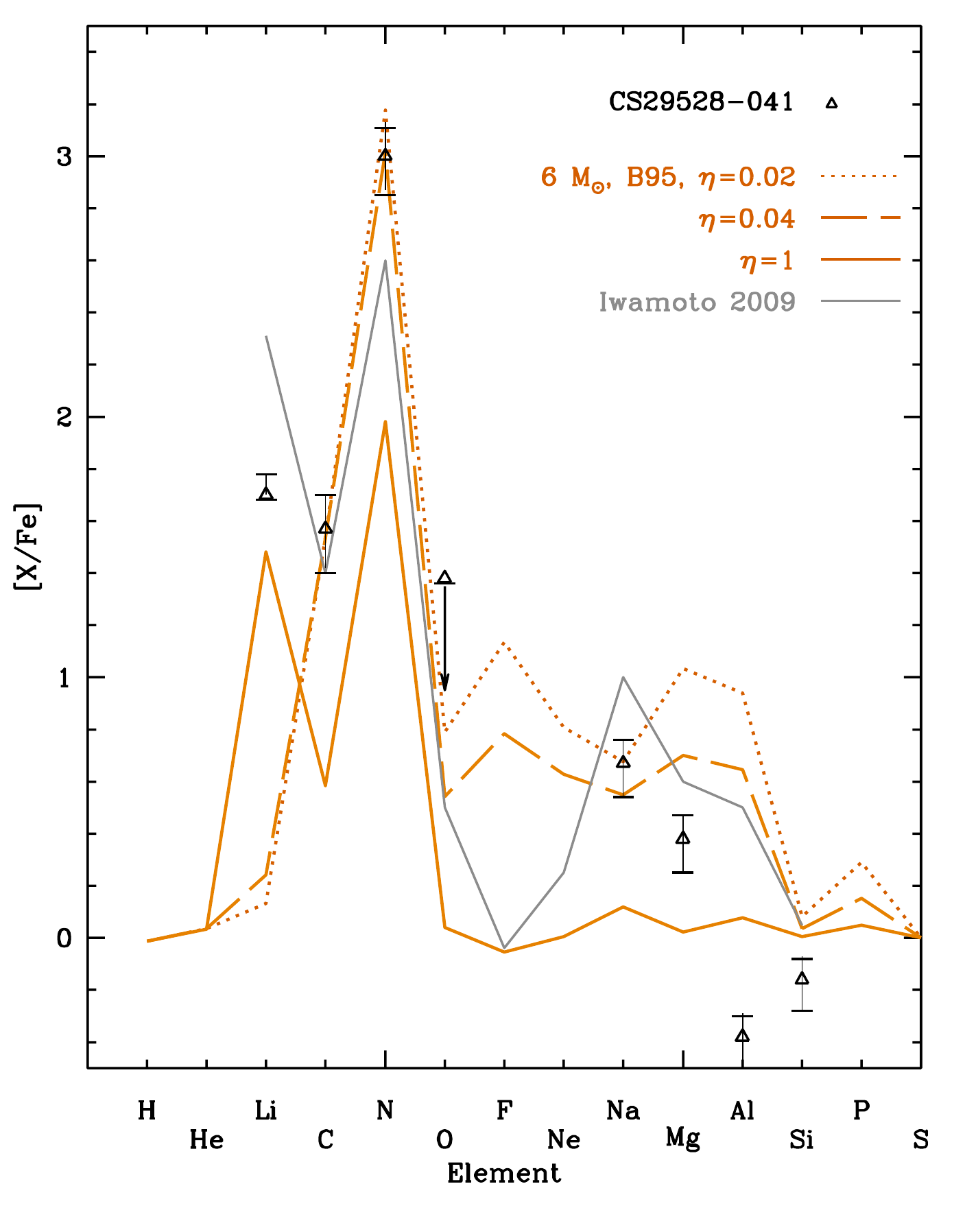}
\caption{Abundance pattern in terms of the ejecta of the 6 \msun{} models
	computed with the Blo95 wind prescription, and assuming dilution with a
	low-mass star envelope (see Figure \ref{fig:panel_obs} and main text for
	details), with $\eta=0.02$,  (dotted line), $\eta=0.04$ (dashed line)
	and $\eta=1$ (solid line). Abundances from the Z=2$\times$10$^{-5}$, 6
	\msun{} model by \citet{iwamoto2009} are shown in grey solid line.
	Symbols represent the observed abundances of CS29528-041. }
\label{fig:best}
\end{center}
\end{figure}

Low-mass EMP stars have a very thin convective envelope, but hydrodynamical
processes such as thermohaline mixing (see, e.g. \citealt{kippenhahn1980},
\citealt{chen2004}, \citealt{stancliffe2007b}),
 can induce further dilution of the accreted material in the (S)AGB companion.
 In particular, \citet{stancliffe2007b} proposed that the accreted material
 could be mixed throughout 90$\%$ of the low-mass accretor. Note however that
 the efficiency of this mechanism was questioned by \citet{aok08}, who observed
 that CEMP turn-off and red giants present very different [C/H] abundances.
 According to these authors, it suggests the absence of efficient mixing prior
 to the red giant phase.   \citet{denissenkov2008} considered the observed
 variations of [C/H] and [N/H] after the first dredge-up of CEMP stars and
 showed that mixing over the top 0.2 \msun{} leads to an overall good agreement
 with observations. The theoretical explanation for the loss of efficiency of
 thermohaline mixing might be related to the effect of gravitational settling
 \citep{stancliffe2009}.

Following this idea, we calculated new abundance patterns associated to the
dilution of 1$\%$ of our ejecta in the uppermost 0.2\msun{} of a low-mass EMP
star of the same initial metallicity. These new abundances are shown in Table
\ref{tab:abuns} (see Appendix \ref{sec:app1}).  For comparison, we also display
in Figure \ref{fig:panel_obs} the expected abundance pattern in terms of the
ejecta from our diluted models, together with the observed abundances of the
eight main sequence N-rich CEMP-s stars of the Galactic Halo which host the
highest surface N-abundances (note that our model stars produce large amounts of
this element), with metallicity in the range $\rm -3.6 < [Fe/H] < -2.9$, that
is, within a 0.3 dex interval about our initial composition ([Fe/H] =-3.2, which
corresponds to $Z=10^{-5}$). 
Abundances were taken from the SAGA database (\citealt{suda2008},
\citealt{suda2011}, \citealt{yamada2013} and \citealt{suda2017a}), which refer
to the following publications for each individual object: \citet{sivarani2006}
for CS29258-041, \citet{behara2010} for SDSSJ1349-0229, \citet{spite2013} for
SDSSJ1114+1828 and SDSSJ1143+2020, \citet{cohen2013} for HE1439-1420, and
\citet{hansen2015} for HE0450-4902 and HE1029-0546. Note that, in general,
agreement between model and observations improves when the wind prescription by
Blo95 is used, with the possible exception of the \iso{12}C/\iso{13}C ratio. 

The high N and C-abundances of CS29258-041 suggests a progenitor undergoing both
HBB and TDU, and thus it is the object whose abundances might be better
explained using intermediate-mass models like ours. \citet{sivarani2006}
themselves compared their observational data with the yields from a 6 \msun{}
model of metallicity [Fe/H]=-2.3 computed by \citet{herwig2004}. This model
gives [C/Fe]=1.2 and [N/Fe]=2.3. By comparison, our 6 \msun{} model computed
with the Blo95 ($\eta =0.02$)  mass-loss rates yields higher CN abundances with
[C/Fe]=2.23 and [N/Fe]=3.88 (see Table \ref{tab:abuns} in Appendix
\ref{sec:app1}). When the parameter $\eta = 1$ is used along with Blo95, we
obtain [C/Fe]=1.17 and [N/Fe]=2.67.
Note also that the initial metallicity of the model by \citet{herwig2004} was
Z=0.0001, so their comparison is rather artificial.  We show in Figure
\ref{fig:best} our best fit to the observed objects of Figure
\ref{fig:panel_obs}.
For this comparison, we used our 6 \msun{} model calculated with the wind
prescription by Blo95 (and $\eta$=0.02), under the assumption of dilution with a
low-mass star envelope.  We also show the results by \citet{iwamoto2009}, who
also compared the yield abundances of his Z=2$\times$10$^{-5}$ 6 \msun{} model,
computed with Blo95 and $\eta=0.1$ to the observed abundances of CS29258-041. 
The match is reasonably good for the abundances of C, N and O (although only an
upper limit for the latter was given). Our Na abundance also appears within the
error bar limits, but our model overproduces Mg by approximately 1 dex, Si by
0.5 dex, and especially Al, by almost 2 dex.  We performed additional
calculations with higher mass-loss rates, aiming at shortening the TP-AGB phase,
and thus HBB (and MgAl chain) responsible for the high overproduction of Al. 
Elements of mass equal to or above that of O show abundances practically equal
to solar, and thus still overproduce Al and Si with respect to the observed
abundances of CS329528-041. 

AGB models are not the only option proposed to explain the surface abundances of
CS29528-041. \citet{sivarani2006} also suggested that metal-poor massive stars
with rotationally-induced mixing could reproduce simultaneously high amounts of
N and C \citep{hirschi2006,mey06,chiappini2006}. However, a recent work
\citep{choplin2017} shows that massive rotating models tend to overproduce Na
and Mg, unless one can assume that only the matter above the He-rich shell is
ejected during the supernova explosion. This hypothesis would be analogous to
the underlying assumptions of the faint supernovae scenario \citep{umeda2005b},
and would work similarly to the wind-only enrichment proposed by \citet{mey06}
and \citet{hirschi2007}. As \cite{choplin2017} point out, a possible way to
distinguish AGB and massive rotating stars scenarios for the formation of EMP
stars could involve the analysis of the ratio of light to heavy s-elements,
which tend to be negative for AGBs \citep{abate2015a}, and positive for massive
rotating stars \citep{cescutti2013}. 

We also explored the possibility that our extremely metal-poor models could
match observations of stars from ultra-faint dwarf galaxies, as they are the
most metal-poor objects known and, as such, are expected to be the least evolved
\citep{simon2019}. The surface N-abundances of these objects were always below
[N/Fe]=1.5.($\pm$0.5), and therefore we could not expect a reasonable match with
our models, in which HBB plays such an important role. Thus additional
comparative analysis is not given, and we defer further considerations till the
sample of observations of objects of ultra-faint dwarf galaxies increases, or
until binary model yields (which could remove HBB) are calculated. 

\section{Summary and discussion} \label{sec:finale}

We have presented the evolution and nucleosynthesis of stellar models with ${\rm
Z=10^{-5}}$ and masses between 3 and 7.5 \msun{}, and compared them to models
computed with similar versions of the Monash stellar evolution code
(\textsc{monstar}).  Determining the trends of relevant evolutionary parameters
with metallicity was not straightforward, due to the lack of fully consistent
input physics sequences amongst the models
(e.g. variations in mass-loss rates, determination of convective boundaries and
low-temperature opacities).  In terms of evolution, the most remarkable trend
with decreasing Z is the longer duration of the TP-(S)AGB phase which, together
with the shorter interpulse periods, leads to a higher number of thermal pulses. 
The use of different (and very uncertain) wind prescriptions affects these
characteristics even more significantly than metallicity. 

Nucleosynthesis in ${\rm Z=10^{-5}}$ massive AGB and Super-AGB stars is governed
by the  efficiency of both HBB and TDU.  The former is a consequence of the high
temperatures at the BCE, which allow the onset of the NeNa-cycle and the
MgAl-chain. The trend to produce significant yields of \iso{4}He, and \iso{14}N
due to HBB is actually shared by very massive AGB and Super-AGB stars of any
metallicity. Specifically, the high \iso{4}He yields produced by these model
stars makes them good candidates to pollute the intracluster medium and thus, a
key to understand multiple populations of globular clusters.

Even though there is  no consensus on the subject, and the dependency on input
physics is very strong, the TDU efficiency of massive AGB and Super-AGB stars
seems to be maintained for lower metallicities, at least down to $Z\sim
10^{-5}$.
As a consequence of deep TDUs,  all the $Z=10^{-4}$ and $Z=10^{-5}$ models
display positive yields of \iso{12}C and \iso{16}O. In addition, due to the high
temperatures achieved in the HeBS and the BCE, the heavy Mg isotopes also
present positive yields.  As shown by \citet{siess2010} and
\citet{doherty2014b}, ${\rm Z=10^{-4}}$ models of massive AGB and Super-AGB
stars produced large amounts of \iso{13}C and \iso{27}Al. ${\rm Z=10^{-5}}$
stars of analogous masses present even higher production factors for these
isotopes and, unlike their $Z=10^{-4}$ counterparts, produce positive (although
low) yields of \iso{21}Ne, \iso{24}Mg and \iso{32}S. 

The presented yields point to the potential relevance of our models as
contributors of \iso{4}He and \iso{14}N, regardless of the wind prescription
used. If real mass-loss rates are biased to low values, our sequences would also
contribute to \iso{16}O, \iso{22}Ne, heavy Mg isotopes, \iso{27}Al, and
\iso{28}Si, although the production of \iso{16}O and \iso{28}Si is mostly due to
massive stars.  Actually, the yields of all the isotopes mentioned above are
about one order of magnitude higher when computed with VW93 than when computed
with Blo95 (with $\eta=0.02$). These differences would impact Galactic chemical
evolution models.
In terms of isotopic ratios of \iso{12}C/\iso{13}C and \iso{25,26}Mg/\iso{24}Mg,
the effects of the mass-loss rates are also crucial. We suggest that, once
future observations of \iso{25,26}Mg/\iso{24}Mg extend below [Fe/H]=-2.5, they
could provide a useful tool to help constraining mass-loss rates in the EMP
regime. The crucial effects of different wind prescriptions on the stellar
yields presented hampers the detection of a clear trend with metallicity, and
emphasize the importance of using consistent grids of nucleosynthetic yields as
inputs for GCE models. 

Comparison of yields from IM EMPs with nitrogen-rich CEMP-s stars from the Halo
is promising enough to further study.  Our models provide a better match to
observations when the mass-loss rates by Blo95 with $\eta=0.02$, more efficient
than VW93, are used.
The uncertain physics of mixing, possibly affected by magnetic buoyancy
\citep{nucci2014}, gravity waves \citep{denissenkov2003, battino2016}, and
rotation \citep{herwig2005,straniero2015,cristallo2015}, also plays a
determining role in the evolution and yields, and its effects should be
explored. 

The environment where $Z=10^{-5}$ stars formed showed the specific signature of
one or a few individual objects, rather than the mixture of a large number of
stellar yields. Therefore it is important to increase the  number of
observational counterparts, and perform a detailed exploration of extended
nucleosynthesis (including s-process), and of the parameter space of theoretical
models, probably combining both the initial compositions corresponding to the
yields of primordial objects and the yields of our early generation models.
Specifically, the effects of stellar winds, which proved critical for model
results at the considered metallicity range, should be consistently taken into
account.


\begin{acknowledgements}
      Part of this work was supported by the Spanish project \it{PID 2019-109363GB-100}, and by the German
      \it{Deut\-sche For\-schungs\-ge\-mein\-schaft, DFG\/} project
      number Ts~17/2--1. 
      LS is a senior FNRS research associate. We thank the anonymous referee for their useful comments and suggestions. 
\end{acknowledgements}

%
%

\bibliographystyle{aa} 
\bibliography{aa_yields1e-5_ref.bib} 

\begin{appendix}
\onecolumn
\section{Model abundances in terms of ejecta.} \label{sec:app1}

The abundance patterns of selected elements in terms of [X/Fe], which may facilitate comparison with observational data and with other theoretical calculations, are presented in table \ref{tab:abuns}. Note that the results corresponding to our actual yields were presented in Figure \ref{fig:isotopes}, and those corresponding to the hypothetical cases in which 1$\%$ of the matter ejected by our models was homogeneously mixed in the 0.2 \msun envelope of an unevolved $Z=10^{-5}$ star, were presented in Figures \ref{fig:panel_obs} and \ref{fig:best}.
 
 \begin{table*}[h]
    \centering
    \caption{Abundances pattern of selected elements in terms of [X/Fe] as given by the ejecta, or under the assumption that 1$\%$ of this matter was homogeneously diluted in the surface 0.2 \msun{} of an unevolved $Z=10^{-5}$ star. Lithium abundance is shown as $<log_{10}(^{7}Li)>=Log_{10}(N(Li)/N(H))+12$. Results using the \citet{vassiliadis1993} and \citet{bloecker1995} with the indicated $\eta$ values are shown.}
\begin{tabular}{l c c c c c c c c c c c c c}
$M_\mathrm{ini}/$\msun & $<log_{10}(^{7}Li)>$ & C & N & O & F & Ne & Na & Mg & Al & Si & P & S \\[1pt]
\hline\\
\multicolumn{13}{c}{\textbf{VW93}}\\
3.0  & -0.49 &  3.19 &  4.97 &  2.30 &  3.74 &  3.85 &  3.95 &  3.73 &  3.35 &  2.15 &  2.62 &  0.29 \\
3.0-dil & -0.04 &  2.14 &  3.91 &  1.27 &  2.69 &  2.80 &  2.90 &  2.68 &  2.30 &  1.13 &  1.57 &  0.02 \\
4.0 & -0.13 &  3.14 &  4.91 &  2.36 &  2.95 &  3.48 &  3.66 &  3.77 &  3.56 &  2.28 &  3.06 &  0.65 \\
4.0-dil & -0.03 &  2.07 &  3.84 &  1.31 &  1.89 &  2.42 &  2.59 &  2.71 &  2.49 &  1.24 &  2.00 &  0.11 \\
5.0  & -0.44 &  2.98 &  4.85 &  2.20 &  2.66 &  3.31 &  3.41 &  3.61 &  3.70 &  2.24 &  3.03 &  0.63 \\
5.0-dil & -0.05 &  2.18 &  4.04 &  1.41 &  1.86 &  2.51 &  2.61 &  2.81 &  2.90 &  1.45 &  2.23 &  0.18 \\
6.0  & -0.46 &  2.78 &  4.75 &  2.04 &  1.93 &  2.96 &  2.92 &  3.39 &  3.64 &  2.04 &  2.76 &  0.43 \\
6.0-ms & -0.06 &  2.07 &  4.03 &  1.34 &  1.24 &  2.25 &  2.21 &  2.67 &  2.92 &  1.34 &  2.05 &  0.12 \\ 
7.0  & -0.49 &  2.69 &  4.54 &  1.87 &  1.46 &  2.49 &  2.36 &  3.07 &  3.35 &  1.95 &  2.55 &  0.34 \\
7.0-dil & -0.07 &  2.04 &  3.88 &  1.23 &  0.85 &  1.84 &  1.71 &  2.42 &  2.69 &  1.31 &  1.90 &  0.10 \\[1pt]
\hline\\
\multicolumn{13}{c}{\textbf{Blo95}}\\
 3.0 &  2.58 &  3.61 &  3.67 &  1.61 &  3.37 &  2.51 &  2.36 &  2.28 &  1.56 &  0.57 &  0.88 &  5.6$\times 10^{-3}$ \\
 3.0-dil &  1.55 &  2.59 &  2.65 &  0.68 &  2.35 &  1.50 &  1.36 &  1.28 &  0.64 &  0.10 &  0.21 & 5.4$\times 10^{-3}$ \\
 4.0 &  1.50 &  2.72 &  4.18 &  1.54 &  2.46 &  2.26 &  2.26 &  2.20 &  1.68 &  0.55 &  1.03 &  9.2$\times 10^{-3}$ \\
 4.0-dil &  0.68 &  1.85 &  3.30 &  0.74 &  1.59 &  1.40 &  1.39 &  1.34 &  0.86 &  0.13 &  0.36 &  9.2$\times 10^{-3}$ \\
 5.0 &  0.94 &  2.42 &  3.99 &  1.30 &  1.77 &  1.90 &  1.91 &  1.92 &  1.64 &  0.38 &  0.94 &  8.0$\times 10^{-3}$ \\
 5.0-dil &  0.32 &  1.65 &  3.22 &  0.62 &  1.03 &  1.15 &  1.16 &  1.17 &  0.91 &  0.09 &  0.36 &  1.4$\times 10^{-3}$ \\
6.0 $(\eta=0.02)$ &  0.49 &  2.23 &  3.88 &  1.43 &  1.81 &  1.45 &  1.29 &  1.70 &  1.60 &  0.31 &  0.76 &  5.8$\times 10^{-3}$ \\
6.0-dil $(\eta=0.02)$  &  0.12 &  1.54 &  3.18 &  0.79 &  1.14 &  0.81 &  0.68 &  1.03 &  0.94 &  0.08 &  0.29 &  1.2$\times 10^{-3}$ \\
6.0 $(\eta=0.04)$   &  0.67 &  2.22 &  3.71 &  1.13 &  1.42 &  1.24 &  1.14 &  1.32 &  1.26 &  0.16 &  0.49 &  2.1$\times 10^{-3}$ \\
6.0-dil $(\eta=0.04)$ &  0.23 &  1.53 &  3.02 &  0.54 &  0.78 &  0.63 &  0.55 &  0.70 &  0.64 &  0.04 &  0.15 &  4.2$\times 10^{-4}$ \\
6.0 $(\eta=1)$   &  2.16 &  1.17 &  2.67 &  0.17 & -0.37 &  0.02 &  0.41 &  0.10 &  0.29 &  0.03 &  0.20 &  4.1$\times 10^{-5}$ \\
6.0-dil $(\eta=1)$ &  1.47 &  0.58 &  1.98 &  0.04 & -0.05 &  0.00 &  0.12 &  0.02 &  0.08 &  0.01 &  0.05 &  8.4$\times 10^{-6}$ \\
7.0     & -0.32 &  2.41 &  3.49 &  1.00 &  1.00 &  1.00 &  0.94 &  1.37 &  1.14 &  0.50 &  0.59 &  6.9$\times 10^{-3}$  \\
7.0-dil & -0.07 &  1.77 &  2.85 &  0.49 &  0.49 &  0.49 &  0.44 &  0.79 &  0.59 &  0.17 &  0.22 &  1.6$\times 10^{-3}$  \\
7.5   & -0.55 &  1.08 &  3.79 & -0.46 & -0.80 & -0.11 & -0.97 & -1.00 & -0.90 &  0.49 & -0.31 &  7.9$\times 10^{-3}$ \\
7.5-dil & -0.09 &  0.56 &  3.17 & -0.07 & -0.10 & -0.02 & -0.11 & -0.11 & -0.10 &  0.18 & -0.06 &  1.9$\times 10^{-3}$ \\
\multicolumn{1}{c}{}
\end{tabular}
\label{tab:abuns}
\end{table*}

\end{appendix}
\end{document}